\documentclass[aps,prd,eqsecnum,showpacs,nofootinbib]{revtex4}

\usepackage[dvipdfmx]{graphicx}
\usepackage{amsmath,amsfonts,amssymb,color,latexsym,theorem,mathrsfs,comment,color}

\newcommand{\ma}[1]{\mbox{$\mathcal{#1}$}}

\newcommand{\D}{{\rm d}}

\newcommand{\we}{\wedge}
\newcommand{\red}[1]{{\textcolor{red}{#1}}}

\begin{document}

\title{
New family of C-metrics in ${\cal N}=2$ gauged supergravity
}

\author{Masato Nozawa${}^1$ and Takashi Torii${}^2$}

\address{
${}^1$General Education, Faculty of Engineering, 
Osaka Institute of Technology, Osaka City, Osaka 535-8585, Japan\\
${}^2$Department of System Design, Osaka Institute of Technology, Osaka City, Osaka 530-8568, Japan
}

\date{\today}

\begin{abstract} 
We present a new family of charged C-metrics in ${\cal N}=2$ gauged supergravity 
in four dimensions. The double Wick rotation of the C-metric allows us to bring our solution 
into a different family of the C-metrics previously found by L\"u and V\'azquez-Poritz. 
In the case of zero acceleration limit, our solution with vanishing charges reduces to the 
scalar haired black holes in AdS with regular horizons. Nevertheless, it turns out that each family of neutral solutions 
fails to veil the curvature singularity by the event horizon, showing that neither of them represents 
the accelerated black holes with a scalar hair.  Physical  solutions without visible curvature singularities are obtained only in the case of nonvanishing charges. 
Causal structures of the solution are spelled out in detail. We also present conditions under which the solution
preserves supersymmetry.

\end{abstract}


\maketitle


\section{Introduction}

Black hole solutions in anti-de Sitter (AdS) space have drawn considerable attention 
from the perspective of holography.  By cause of the admirable nature of duality, 
classical AdS black holes have provided an extremely valuable arena for exploring the strongly coupled 
dual gauge theories and the condensed matter physics applications. 

In the asymptotically flat spacetimes, stationary black holes in vacuum are 
essentially unique and identified entirely by mass and angular momentum 
\cite{Israel:1967wq,MRS,robinson,bunting,Robinson:1975bv,Nozawa:2018kfk}. 
Despite the primary importance of AdS black holes, 
similar classification of black holes in AdS is a much more difficult task to implement. 
There is no known formalism of comparable power for exhaustive classification of AdS black holes, even though 
one centers on the static case. Some progresses for partial classification have been achieved  under fairly restrictive assumptions \cite{Sudarsky:2002mk,Anderson:2002xb}. The construction of exact solutions is likewise unfeasible in a straightforward fashion, since 
the cosmological constant or the scalar field potential destroys the symmetry of 
the reduced target space of nonlinear sigma model, which prevents us to generate new solutions from 
 simpler seed solutions \cite{Klemm:2015uba}. 

In the teeth of the above adversity, 
it turns out that AdS black holes enjoy substantially rich varieties and are endowed with physically  interesting  properties.  
Of most prominence is that some black holes admit scalar `hairs.' Within the framework of supergravity, 
exact solutions describing static AdS black holes with a nontrivial scalar configuration have been constructed in \cite{Anabalon:2012ta,Feng:2013tza}
(see also \cite{Anabalon:2012sn,Anabalon:2013eaa,Anabalon:2020pez,Lu:2013eoa,Lu:2013ura}).\footnote{Many authors have explored scalar haired AdS black holes with diverse potentials. Nonexhaustive list of references 
for canonical scalar field in Einstein's gravity is \cite{Torii:2001pg,Kolyvaris:2010yyf,Gonzalez:2013aca,Mahapatra:2020wym}.}
These solutions embody the manifestation of non-uniqueness, since the theory under study obviously possesses the Schwarzschild-AdS black hole for which the scalar field is frozen to a constant. 
A more unanticipated and tantalizing facet is that these scalar haired black holes themselves are not unique. 
Ref. \cite{Faedo:2015jqa} has presented a new scalar haired black hole in the same theory 
as in \cite{Anabalon:2012ta,Feng:2013tza}. 
These works prompt the questions with regard to the diversity of AdS black holes.

In the present paper, we undertake this problem by focusing on the 
accelerating black holes broadly termed the C-metric. 
The vacuum C-metric \cite{LeviCivita,Newman1961,Ehlers:1962zz} describes 
a pair of black holes undergoing uniform acceleration in opposite directions \cite{Kinnersley:1970zw}.  
The acceleration of black holes is provided by a conical 
deficit angle corresponding to the cosmic string extending out to infinity, or a strut with a negative tension 
which stretches between two black holes. Specifically, the C-metric can be realized as a perturbation 
of the Schwarzschild black hole with a distributional stringy source \cite{Kodama:2008wf}. 
The vacuum C-metric falls into Petrov type-D and Weyl class of solutions \cite{Bonnor}. 
Causal structures and physical properties of the vacuum C-metric have been fully
investigated in \cite{Hong:2003gx,Letelier:1998rx,Griffiths:2006tk,Lim:2014qra}. 
The C-metric in AdS has been also studied intensively from various points of view: 
causal structures 
\cite{Podolsky:2002nk,Dias:2002mi,Krtous:2005ej}, thermodynamics \cite{Appels:2016uha, Appels:2017xoe,Astorino:2016ybm,Zhang:2018hms,Wang:2022hzh},
minimal surfaces \cite{Xu:2017nut} and 
quasi-normal modes \cite{Nozawa:2008wf,Destounis:2020pjk,Destounis:2022rpk}.
Here we consider the supergravity generalization of the AdS C-metric, for which
the AdS vacuum is realized by the scalar field potential rather than a pure cosmological constant.

A charged C-metric in $\ma N=2$ supergravity has been discovered in \cite{Lu:2014ida,Lu:2014sza}. 
The bosonic Lagrangian studied in \cite{Lu:2014ida,Lu:2014sza} consists of Einstein-Maxwell-dilaton theory with an arbitrary coupling constant and is identical to the cases considered in \cite{Anabalon:2012ta,Feng:2013tza}. 
The C-metric solution in \cite{Lu:2014ida,Lu:2014sza} reduces in the zero acceleration limit
to the spherical solution in \cite{Anabalon:2012ta,Feng:2013tza}. 
It is worth mentioning that the spherically symmetric solution in \cite{Anabalon:2012ta,Feng:2013tza}  describes a naked singularity 
instead of a black hole in the neutral case, whereas 
the spherical solution in \cite{Faedo:2015jqa} allows a parameter range in which the event horizon exists. 
It is then pertinent to deduce that the neutral C-metric in \cite{Lu:2014ida,Lu:2014sza} would represent 
a pair of accelerated naked particles and there should exist another family of `hairy C-metrics' which incorporates the spherical solution in \cite{Faedo:2015jqa}. 
This is a  prime motivation of the present paper. 

Bearing these prospects in mind, we present a new charged C-metric in $\ma N=2$ supergravity. 
We demonstrate that our solution is converted to the one in \cite{Lu:2014ida,Lu:2014sza} via 
double Wick rotation. This transformation is missing in the zero acceleration limit and 
properly accounts  for the existence of two distinct hairy solutions within the same theory. 
The emphasis of the present article is placed on the causal structure of the C-metric solution.
In the neutral limit, our solution displays some peculiarities, 
most notably the event horizon disappears in any range of parameters. 
This comes as a surprise since the neutral solution reduces in the vanishing acceleration case
to the hairy black hole solution in \cite{Faedo:2015jqa}. 
This proclaims that the zero acceleration limit of the solution is discontinuous. 
Nevertheless, the charged solution can have a parameter range 
in which the horizon exists. 

We organize the present article as follows. 
In the next section, we give a quick review of ${\cal N}=2$ supergravity 
with Abelian Fayet-Iliopoulos gaugings. Upon truncation, we will see that the bosonic 
theory reduces to the Einstein-Maxwell-dilaton gravity with a potential which is expressed in 
terms of a real superpotential. 
Section \ref{sec:Cmetric} provides our new C-metric and discusses
various limits of the solution. Causal structures and physical properties of the solution
are examined in section \ref{sec:property}. 
Section \ref{summary} concludes our paper with short summary and future outlooks. 
Supersymmetry of the C-metric will be investigated in appendix. 
We employ the units $c=8\pi G=1$ throughout the paper.

\section{Fayet-Iliopoulos gauged supergravity}

Let us consider the $\ma N=2$ gauged supergravity coupled to $n_V$ number of 
abelian vector multiplets in four dimensions~\cite{Andrianopoli:1996cm} (see, e.g, 
\cite{Freedman:2012zz,Trigiante:2016mnt, DallAgata:2021uvl} for recent reviews). 
We follow the conventions of \cite{Cacciatori:2008ek}. 
The bosonic field contents consist of the vectors $A^I{}_\mu$ ($I=0, 1, ..., n_V$)
and the complex scalars $z^\alpha$ ($\alpha =1, ..., n_V$). 
These scalars parametrize the special K\"ahler manifold corresponding to the 
$n_V$-dimensional Hodge-K\"ahler manifold endowed with a symplectic bundle. 
The symplectic bundle is characterized by a covariantly holomorphic section
\begin{align}
\label{}
\ma V= \left(
\begin{array}{c}
X^I    \\
F_I   
\end{array}
\right)\,, \qquad 
\ma D_{\bar \alpha}\ma V= \partial_{\bar \alpha}\ma V-\frac 12 (\partial _{\bar \alpha}\ma K)\ma V =0 \,, 
\end{align}
where $\ma K=\ma K(z^\alpha, \bar z^\alpha)$ is the K\"ahler potential and 
$\ma D_\alpha$ denotes the K\"ahler covariant derivative. 
The covariantly holomorphic section obeys the following symplectic constraint
\begin{align}
\label{}
\langle \ma V, \bar{\ma V}\rangle \equiv X^I\bar F_I-F_I\bar X^I= i \,, \qquad 
\langle \ma V, \partial_\alpha {\ma V}\rangle=0 \,, 
\end{align}
where $\langle~ , ~\rangle $ stands for the symplectic inner product induced by 
the symplectic metric $\Omega=i\sigma_2 \otimes I_{n_V}$. 
Writing
\begin{align}
\label{}
\ma V= e^{K/2}v\,, \qquad 
v= \left(
\begin{array}{c}
Z^I    \\
Y_I   
\end{array}
\right), 
\end{align}
$v$ denotes the symplectic section
$\partial _{\bar \alpha} v=0$. Assuming the invertibility of the matrix $(X^I~\partial_\alpha X^I)$, 
the symplectic constraint implies the existence of a 
prepotential $F$ satisfying 
\begin{align}
\label{}
Y_I=\frac{\partial}{\partial Z^I}F(Z)\,, \qquad 
F(\lambda Z)=\lambda ^2 F(Z) \,. 
\end{align}
Throughout the paper, we assume the existence of the prepotential. 

The coupling between the scalars
$z^{\alpha}$ and the vectors $A^I{}_{\mu}$ is controlled by the 
complex matrix ${\cal N}_{IJ}$ which  is defined by the relations
\begin{align}
\label{NIJ}
F_I = {\cal N}_{IJ}X^J\,, \qquad {\cal D}_{\bar\alpha}\bar F_I = {\cal N}_{IJ}
{\cal D}_{\bar\alpha}\bar X^J\,. 
\end{align}
Then, the bosonic Lagrangian reads
\begin{align}
\label{Lag0}
\ma L=\frac 12 (R-2 V) \star 1 -g_{\alpha\bar \beta}\D z^\alpha \we \star \D \bar z^{\bar\beta} 
+\frac 12 I_{IJ}F^I \we \star F^J +\frac 12 R_{IJ}F^I \we  F^J \,.
\end{align}
where we have written $I_{IJ}={\rm Im}\ma N_{IJ}$, $R_{IJ}={\rm Re}\ma N_{IJ}$
and $F^I=\D A^I$ is the electromagnetic field strength. 
The scalar potential is 
\begin{align}
\label{}
V = -2g_Ig_J\left(I^{IJ}+8\bar X^IX^J\right)\,, 
\label{pot}
\end{align}
where $I^{IJ}$ is the inverse of $I_{IJ}$ and 
$g_I$ denote the Fayet-Iliopoulos coupling constants. 
In what follows, we assume $g_I>0$.

Einstein's equations derived from the Lagrangian (\ref{Lag0}) read
\begin{align}
\label{}
R_{\mu\nu}-\frac 12 R g_{\mu\nu}=&\, T_{\mu\nu}\,, 
\end{align}
where 
\begin{align}
\label{}
T_{\mu\nu}&=-I_{IJ} \left(F^I_{\mu\rho}F_\nu^{J\rho}-\frac 14
g_{\mu\nu}F^I_{\rho\sigma}F^{J\rho\sigma} 
\right) +2g_{\alpha\bar \beta}\left(
\nabla_{(\mu} z^\alpha \nabla_{\nu)} \bar z^{\bar \beta} -\frac 12 g_{\mu\nu}
\nabla_\rho z^\alpha \nabla^\rho \bar z^{\bar \beta}
\right) -V g_{\mu\nu} \,. 
\end{align}
The gauge fields obey 
\begin{align}
\label{}
\D \left(I_{IJ}\star F^J +R_{IJ}F^J\right)= &\, 0 \,. 
\end{align}
Lastly, the scalar field equations boil down to
\begin{align}
\label{}
 \nabla^2 z^\alpha+{}^{T\!} \Gamma^\alpha{}_{\beta\gamma}\nabla_\mu z^\beta \nabla^\mu z^{\gamma}-g^{\alpha\bar \beta}\partial_{\bar \beta}V
+\frac 14g^{\alpha\bar\beta}\left[(\partial_{\bar \beta}I_{IJ})F^I_{\rho\sigma}F^{J\rho\sigma}
- (\partial_{\bar \beta}R_{IJ})F^I_{\rho\sigma}\star F^{J\rho\sigma}\right]=&\,0
\label{scalareq}
\end{align}
where ${}^{T\!} \Gamma^\alpha{}_{\beta\gamma}$ is the  affine connection of the target space.

\subsection{Model}

We focus on a one-parameter family of ${\cal N}=2$ supergravity models in which 
the prepotential is given by \cite{Faedo:2015jqa}
\begin{align}
\label{prepotential}
F(X)=-\frac i4 (X^0)^n (X^1)^{2-n} \,,
\end{align}
corresponding to $n_V=1$ involving a single complex scalar. 
For the special choice of the parameter $n=1, 1/2$ and $3/2$, 
the theory is obtained by the truncation of what is called the STU model and is 
embedded into the eleven dimensional supergravity \cite{Cvetic:1999xp,Azizi:2016noi}. 
The prepotential of the STU model is 
\begin{align}
\label{}
F_{\rm STU}(X)=-\frac i4\sqrt{X^0X^1X^2X^3}\,,
\end{align}
for which the three complex scalars $z^i=X^i/X^0$  parametrize the coset $[{\rm SL}(2,\mathbb R)/{\rm SO}(2)]^3$.
The $n=1$ is obtained when a single scalar field is turned on $z^2=z^3=0$, while 
the $n=1/2, 3/2$ cases correspond to the diagonal truncation $z^1=z^2=z^3=z$ \cite{Duff:1999gh}.

Setting $Z^0=1$ and $Z^1=z$, 
the symplectic vector reads
\begin{align}
\label{}
v = \left(\begin{array}{c}
1	\\
z	\\
-\dfrac{i}{4} \, n z^{2-n}	\\
-\dfrac{i}{4} \, (2-n) z^{1-n}
\end{array}\right)\,,
\end{align}
and  the K\"ahler potential is given by
\begin{align}
\label{}
e^{-\mathcal{K}} = \frac{1}{4} \left[n(z^{2-n} + \bar{z}^{2-n}) + (2-n)(z^{1-n}\bar{z} + z\bar{z}^{1-n})
\right]\,.
\end{align}
When $n=1$, $1/2$ and $3/2$, the scalar manifold corresponds to the 
coset ${\rm SU}(1,1)/{\rm U}(1)$.

To proceed, we would like to further truncate the theory to the real scalar $z=\bar z$. 
This is possible if ${\rm Im}\,z=0$ is consistent with the equations of motion (\ref{scalareq}).  
After some computations, one can ascertain that this is indeed the case, as far as we 
concentrate on the purely electrically or magnetically charged solutions
\begin{align}
\label{truncation}
F^I \we F^J=0 \,.  
\end{align} 
In this truncated case,  the condition ${\rm Im}\ma N_{IJ}<0$ requires $0<n<2$ and 
the bosonic Lagrangian is simplified to 
\begin{align}
\label{Lag0s}
\ma L=\frac 12 (R-2V)\star 1-\frac 12 \D \Phi \we \star \D \Phi 
-\frac 18 ne^{\pm\sqrt{\frac{2(2-n)}{n}}\Phi}F^0 \we \star F^0
-\frac 18 (2-n) e^{\mp\sqrt{\frac{2n}{2-n}}\Phi}F^1 \we \star F^1
\end{align}
where we have set $z=\exp ({\pm\sqrt{2/[n(2-n)]}\Phi})$. 
The scalar potential $V$ is expressed in terms of the 
real superpotential $W$ as
\begin{align}
\label{pot0}
V=4 \left[2 (\partial _\Phi W)^2-3 W^2 \right]\,, 
\end{align}
where 
\begin{align}
\label{W0}
W(\Phi)=g_IX^I=g_0 e^{\mp\sqrt{\frac{2-n}{2n}}\Phi}+g_1 e^{\pm\sqrt{\frac{n}{2(2-n)}}\Phi} \,. 
\end{align}
The present theory (\ref{Lag0s}) has the following symmetry 
\begin{align}
\label{sym}
n\leftrightarrow 2-n \,, \qquad 
\Phi \leftrightarrow -\Phi \,, \qquad 
g_0 \leftrightarrow g_1 \,, \qquad 
F^0 \leftrightarrow F^1\,,
\end{align}
corresponding to the 
interchange of $X^0$ and $X^1$. 

The scalar potential (\ref{pot0}) admits at most two critical points
\begin{align}
\label{}
\Phi_0=\pm\sqrt{\frac{n(2-n)}{2}}\log\left(\frac{g_0(2-n)}{g_1 n}\right)\,, \qquad 
\Phi_1=\pm\sqrt{\frac{n(2-n)}{2}}\log\left(\frac{g_0(2-n)(1-2n)}{g_1 n(3-2n)}\right)\,. 
\end{align}
Under the present proviso $g_I>0$, the critical point $\Phi_0$ exists all values of $0<n<2$, 
while  the critical point $\Phi_1$ is absent for $1/2 \le n \le 3/2$. 
Both of these critical points correspond to the AdS vacua. 
The former critical point $\Phi_0$ also extremizes  the superpotential (\ref{W0}), i.e., 
this is a supersymmetric AdS vacuum. 
At $\Phi=\Phi_0$, we have
\begin{align}
\label{}
V=-3g^2 \,, \qquad 
\partial_\Phi^2 V=-2 g^2 \,, 
\end{align}
where $g $ denotes the reciprocal of the AdS radius given by 
\begin{align}
\label{}
g\equiv 4\sqrt{\Big(\frac{g_0}{n}\Big)^n \Big( \frac{g_1}{2-n}\Big)^{2-n}} \,. 
\end{align}
Notably, the mass square $m^2=\partial_\Phi^2 V$ lies in the unitary range 
$m_{\rm BF}^2 <m^2 <m_{\rm BF}^2+g^2$, where $m_{\rm BF}^2=-9g^2/4$ is 
the Breitenlohner-Freedman bound \cite{Breitenlohner:1982jf}. 
When the mass parameter of the AdS extremum lies in this characteristic range, 
the scalar field may be subjected to the `mixed' boundary conditions.
In this case, the slower fall-off mode of the scalar field also survives and 
back reacts nontrivially on the metric.\footnote{Denoting the conformal dimensions of the scalar field as $\Delta_\pm =(3\pm \sqrt{4m^2g^{-2} +9})/2$, the scalar field behaves as $\Phi\sim \Phi_-/r^{\Delta_-}+\Phi_+/r^{\Delta_+}$ around AdS boundary ($r\to \infty$). When $ m^2\ge  m_{\rm BF}^2+g^2$, we must impose the Dirichlet boundary condition $\Phi_-=0$ since the slower fall-off mode $\Phi_-$ is not normalizable. When $m_{\rm BF}^2\le m^2\le m_{\rm BF}^2+g^2$ which occurs in the present case, both modes  are normalizable and the slower fall-off mode $\Phi_-$ might be nonvanishing.  
See \cite{Ishibashi:2004wx,Hertog:2004dr,Henneaux:2006hk} for details.} 
Despite the apparent divergence of 
conventional charges, one can still find generators of asymptotic symmetries 
and the corresponding charges of finite value.

To simplify the system further,  let us relabel 
\begin{align}
\label{}
&\phi=\, \sqrt 2(\Phi-\Phi_0) \,, \qquad 
\alpha=\mp \sqrt{\frac{2-n}{n}} \in \mathbb R\,, 
\notag \\
& (F^0, F^1) \to \left(\frac{g_0\alpha^2}{g_1}\right)^{\frac 1{1+\alpha^2}}
\frac{2\sqrt{1+\alpha^2}}{\alpha}\left(\frac{g_1}{g_0\alpha}F^0,  F^1\right)
\end{align}
for which 
\begin{align}
\label{V}
V(\phi)=4 \left[4(\partial _\phi W)^2-3 W^2 \right]\,, \qquad 
W(\phi)=\frac{g}{2(1+\alpha^2)}\left(e^{\frac{\alpha }{2}\phi}+\alpha^2 e^{-\frac{\phi}{2 \alpha}}\right)\,. 
\end{align}
The Lagrangian is then reduced to
\begin{align}
\label{Lag}
\ma L=\frac 12 (R-2V(\phi))\star 1-\frac 14 \D \phi \we \star \D \phi 
-e^{-\alpha \phi}F^0 \we \star F^0
-e^{\frac{1}\alpha \phi}F^1 \we \star F^1 \,.
\end{align}
The symmetry (\ref{sym}) now amounts to 
\begin{align}
\label{sym2}
\alpha \leftrightarrow  -\frac 1\alpha\,, \qquad 
F^0 \leftrightarrow F^1\,.
\end{align}
On top of this, 
the above Lagrangian (\ref{Lag}) admits a trivial symmetry
\begin{align}
\label{sym3}
\alpha \leftrightarrow  -\alpha\,, \qquad \phi \leftrightarrow -\phi \,.
\end{align}
This enables us to focus on the domain $ \alpha >0$, which we assume hereafter.

The values 
$\alpha=1$, $\sqrt 3$ and $1/\sqrt 3$ are special since theories with these special parameters can be embedded into the maximal ${\cal N}=8$ gauged supergravity. 

\section{New C-metric solution}
\label{sec:Cmetric}

A new gravitational solution for the system (\ref{Lag}) with (\ref{V}) is
\begin{align}
\label{}
\D s^2=&\,\frac 1{A^2(x-y)^2} \left[
h(x)^{\frac{2\alpha^2}{1+\alpha^2}}\left(-h(y)^{\frac{1-\alpha^2}{1+\alpha^2}}\Delta_y(y)\D t^2+\frac{\D y^2}
{h(y)^{\frac{1-\alpha^2}{1+\alpha^2}}\Delta_y(y)}\right)\right.
\notag \\ & \left.+h(y)^{\frac{2\alpha^2}{1+\alpha^2}}
\left(\frac{\D x^2}{h(x)^{\frac{1-\alpha^2}{1+\alpha^2}}\Delta_x(x)}
+h(x)^{\frac{1-\alpha^2}{1+\alpha^2}}\Delta_x(x) \D \varphi^2 
\right)
\right]\,, \label{sol1}\\
 \phi=&-\frac{2\alpha}{1+\alpha^2} \log \left(\frac{h(y)}{h(x)}\right)\,, 
 \qquad 
  A^0 = \frac{q_0x }{\sqrt{1+\alpha^2}} \D \varphi \,, \qquad A^1=\frac{\alpha q_1 x}{\sqrt{1+\alpha^2} h(x)}  \D \varphi \,. 
  \label{sol12}
\end{align}
where 
\begin{align}
\label{}
h(x)=1+A r_0 x \,,
\end{align}
and
\begin{subequations}
\label{general}
\begin{align}
\Delta_y(y)=& \, -a_0 -2 a_1 y-a_2 y^2+\frac{A q^2_0y^3}{r_0}-\frac{Aq_1^2y^3}{r_0 h(y)} \,, \\
\Delta_x(x)=& \, a_0+2a_1 x+a_2 x^2 -\frac{A q^2_0x^3}{r_0}+\frac{Aq_1^2x^3}{r_0 h(x)}+\frac{g^2}{A^2} h(x)^{\frac{3\alpha^2-1}{1+\alpha^2}}\,.
\end{align}
\end{subequations}
Here,  $A$, $r_0$, $q_{0,1}$, $a_{0, 1,2 }$ are arbitrary constants. 
Since both of the gauge fields are magnetic, the condition  (\ref{truncation}) for the consistent truncation
is indeed fulfilled. 

Inasmuch as the hypersurface-orthogonality of Killing vectors $\partial/\partial t$ and $\partial/\partial \varphi$, 
the metric (\ref{sol1}) is static and axially symmetric. The scalar field and the gauge fields are also invariant under 
these symmetries. An elementary computation verifies that the solution (\ref{sol1}) belongs to the Petrov type D. 
These properties are shared by the conventional C-metric in the Einstein-Maxwell-$\Lambda$ system.

The solution (\ref{sol1}) is not manifestly symmetric under (\ref{sym2}). In order to illustrate this symmetry, 
we adopt new variables
\begin{align}
\label{}
x=-\frac{x'}{1+A r_0 x'}\,, \qquad 
y=-\frac{y'}{1+A r_0 y'}\,,
\end{align}
with the property $h(x)=1/h(x')$. 
In terms of these `primed' coordinates, the metric, the scalar field and the gauge fields are indeed form-invariant, 
provided 
\begin{align}
\label{}
q_0'=&\, q_1\,, \qquad 
q_1'=-q_0 \,, \qquad 
a_0 '=a_0 \,, \notag \\
a_1'=&\, -a_1 +A r_0 a_0\,,\qquad 
a_2'= a_2-2A r_0 a_1+A^2 r_0^2 a_0\,.
\label{alphainv:para}
\end{align}
This is no more than the relabeling of parameters, i.e.,  the transformation (\ref{sym2})
maps a  solution into another one within the same family of solutions. 

Remark that some of the seven parameters ($A$, $r_0$, $q_{0,1}$, $a_{0, 1,2 }$) of the solution
are unphysical and gauged away. This becomes evident by noting that 
the solution (\ref{sol1}) admits the following shift and scaling symmetry
\begin{align}
\label{shift}
x=b_0 x''+b_1 \,, \qquad y= b_0 y''+b_1 \,, \qquad 
t=b_2 t''\,, 
\qquad 
\varphi=b_2 \varphi''\,,
\end{align}
together with 
\begin{align}
\label{}
A''=&\delta_1 \sqrt{\frac{b_0}{b_2}}h(b_1)^{\frac{-\alpha^2}{1+\alpha^2}} A\,, 
\qquad  
r_0''=\,\delta_1 \sqrt{b_0 b_2}h(b_1)^{-\frac{1}{1+\alpha^2}} r_0\,,
\notag \\
\qquad 
q''_0=&\delta_2 b_0b_2 q_0 \,,\qquad  
q''_1= \delta_3  \frac{b_0b_2}{h(b_1)^2}q_1 \,,
\notag \\  
a_2''=&\left(a_2-\frac{3Ab_1 q_0^2}{r_0}+
\frac{q_1^2}{r_0^2}\left(1-h(b_1)^{-3}\right)\right) b_0 b_2 h(b_1)^{\frac{1-\alpha^2}{1+\alpha^2}}\,, 
\notag \\  
a_1''=&\left(a_1+a_2 b_1-\frac{3Ab_1^2q_0^2}{2r_0}
+\frac{Ab_1^2q_1^2(1+2h(b_1))}{2r_0 h(b_1)^2}
\right)b_2 h(b_1)^{\frac{1-\alpha^2}{1+\alpha^2}}\,, 
\notag \\
a_0''=& -\Delta_y(b_1)\frac{b_2}{b_0}h(b_1)^{\frac{1-\alpha^2}{1+\alpha^2}}\,. 
\end{align}
Here,  
$b_0$, $b_1$ and $b_2$ are constants and $\delta_{1,2,3}=\pm 1$. Supposed $a_2\ne 0$, this 
three-parameter family of coordinate freedom 
permits us to scale $a_0$ and $a_2$,  and 
take $a_1$ as  any value we wish. Moreover, the appropriate sign choice of $\delta_1=\pm 1$ allows us to choose $A>0$ without loss of generality.

In the following, we would like to view the constant $A$ as an acceleration parameter. 
Unfortunately, 
the present metric (\ref{sol1}) fails to admit the $A \to 0$ limit in the present form, due to the overall factor $A^{-2}$. 
To overcome this difficulty, let us  introduce rescaled coordinates
\begin{align}
\label{}
r=-\frac 1{A y}\,, \qquad \tau  =\frac 1A t \,, 
\end{align}
in terms of which one can recast the metric and the scalar field into 
\begin{align}
\label{}
\D s^2=&\,\frac 1{(1+A r x)^2} \left[
h(x)^{\frac{2\alpha^2}{1+\alpha^2}}\left(-f(r)^{\frac{1-\alpha^2}{1+\alpha^2}}\Delta_r(r)\D \tau^2+\frac{\D r^2}
{f(r)^{\frac{1-\alpha^2}{1+\alpha^2}}\Delta_r(r)}\right)\right.
\notag \\ & \left.+r^2 f(r)^{\frac{2\alpha^2}{1+\alpha^2}}
\left(\frac{\D x^2}{h(x)^{\frac{1-\alpha^2}{1+\alpha^2}}\Delta_x(x)}
+h(x)^{\frac{1-\alpha^2}{1+\alpha^2}}\Delta_x(x) \D \varphi^2 
\right)
\right]\,, \label{sol1rx}\\
 \phi=&-\frac{2\alpha}{1+\alpha^2} \log \left(\frac{f(r)}{h(x)}\right)\,. \label{sol1phi}
\end{align}
The gauge fields are still given by (\ref{sol12}). 
Here, we have defined
\begin{align}
\label{Deltar0}
\Delta_r (r)= \,& -a_2+2 a_1 A r-a_0 A^2 r^2-\frac{q^2_0}{r_0 r}+
\frac{q_1^2}{r_0 r f(r)}\,, \qquad 
f(r)= 1-\frac{r_0}{r} \,.
\end{align}

We are now in a position to discuss the $A\to 0$ limit of (\ref{sol1rx}). 
Nontrivial relations arise only from the structure function $\Delta_x(x)$ around $A=0$ as 
\begin{align}
\label{para:sol1}
a_2=O(A^0)\,, \qquad 
a_1=-\frac{(3\alpha^2-1)g^2 r_0}{2A(1+\alpha^2)}+O(A^0) \,, \qquad 
a_0= -\frac{g^2}{A^2}+O(A^0)\,. 
\end{align}
Defining
\begin{align}
\label{}
k=-a_2-\frac{(\alpha^2-1)(3\alpha^2-1)}{(1+\alpha^2)^2}g^2 r_0^2 \,, 
\end{align}
the coordinate freedom (\ref{shift}) allows us to normalize  $k$ to be $\pm 1$ or $0$
and $O(A^0)$ term in $a_1$ to vanish. 
The last freedom is to scale the $O(A^0)$ term in $a_0$. 
Requiring the metric keeps  the Lorentzian signature in the $g=q_{0,1}=0$ case for any value of $k$, 
one finds that the $O(A^0)$ term in $a_0$ should be scaled to be $+1$, i.e., 
\begin{subequations}
\label{Deltarx}
\begin{align}
\Delta_r (r)= & \, k
-A^2 r^2 -\frac{q_0^2}{r_0 r}
+\frac{q_1^2}{r_0r f(r)}+g^2 \left(r^2 -\frac{3\alpha^2-1}{1+\alpha^2} r_0 r +\frac{(\alpha^2-1)(3\alpha^2-1)}{(1+\alpha^2)^2}r_0^2\right)\,, \label{Deltar} 
\\ 
\Delta_x(x)= &\, 1-k x^2-\frac{Aq_0^2}{r_0}x^3+\frac{Aq_1^2x^3}{r_0 h(x)}
+g^2 \left(
\frac{h(x)^{\frac{3\alpha^2-1}{1+\alpha^2}}-1}{A^2}-\frac{3\alpha^2-1}{(1+\alpha^2)A}r_0 x 
-\frac{(\alpha^2-1)(3\alpha^2-1)}{(1+\alpha^2)^2}r_0^2x^2
\right)\,.\label{Deltax}
\end{align}
\end{subequations}
Bringing back to the original coordinate $y$, the structure function $\Delta_y (y)$ reads
\begin{align}
\label{Deltays}
\Delta_y (y)=&\, -1+k y^2+\frac{Aq_0^2}{r_0}y^3-\frac{Aq_1^2y^3}{r_0 h(y)}+g^2 \left(\frac 1{A^2}+\frac{3\alpha^2-1}{(1+\alpha^2)A}r_0y
+\frac{(\alpha^2-1)(3\alpha^2-1)}{(1+\alpha^2)^2}r_0^2 y^2 
\right) \,. 
\end{align}
Quite surprisingly, the terms proportional to $g^2$ in 
$\Delta_x(x)$ vanish
when $\alpha=1$, $\sqrt 3$ and $1/\sqrt 3$.

It turns out that the solution is characterized by five  parameters $k$, $A$, $r_0$, $q_{0,1}$, while 
  $g$ and $\alpha$ parametrize the theory (\ref{Lag}). 
In the following subsections, we elucidate the physical meaning of the above parameters by
taking various limits of the solution (\ref{sol1}). 
We will see that $k$ controls the topology, $A$ is the acceleration, $r_0$ 
encodes the mass and $q_{0, 1}$ denote magnetic charges.

\subsection{$r_0=q_{0,1} =0$ case: AdS}

If we set $r_0=q_{0,1} =0$, the scalar field and the gauge fields become trivial. The metric is now simplified to 
\begin{align}
\label{AdS}
\D s^2=\frac 1{A^2(x-y)^2}\left(-\Delta_y(y)\D t^2+\frac{\D y^2}{\Delta_y(y)}
+\D \Sigma^2_k (x, \varphi)
\right)\,, 
\end{align}
where $\Delta_y(y)=g^2 A^{-2}-1+k y^2$ 
and 
\begin{align}
\label{2D}
\D \Sigma_k^2(x, \varphi)\equiv \frac{\D x^2}{1-k x^2}
+\left(1-k x^2\right)\D \varphi^2 \,.
\end{align}
The two dimensional metric $\D \Sigma_k^2$ stands for the maximally symmetric space with 
a constant scalar curvature ${}^{2\!} R=2k $. The angular coordinate $\varphi$ has a canonical 
periodicity $2\pi$ for $k=\pm 1$.
Indeed, the above metric satisfies $R_{\mu\nu\rho\sigma}=-2g^2 g_{\mu[\rho}g_{\sigma]\nu}$ and recovers AdS written in the unusual coordinates. 
The above coordinate patch is the analogue of the Rindler coordinate in Minkowski spacetime. To illustrate this, let us consider a static 
observer sitting at $|y|\to \infty$ with constant $x, \varphi$. We see that this observer  
undergoes an acceleration $a^\mu =u^\nu \nabla_\nu u^\mu$ 
with constant magnitude $|a^\mu |=A$.

To demonstrate the explicit coordinate transformation to 
more familiar AdS patches,  we tentatively suppose $g^2>A^2$ and  define new coordinates as
\begin{align}
\label{corrdAdS}
R=\frac{\sqrt{F_0(x, y)}}{\sqrt{g^2-A^2}gA(x-y)}\,, \qquad 
w=\frac{g^2x-A^2(x-y)}{\sqrt{F_0(x,y)}}\,, \qquad 
T=\frac{\sqrt{g^2-A^2}}{Ag}t\,, 
\end{align}
where $F_0(x,y)\equiv g^2[g^2-A^2(1-k y^2)]-A^2k (g^2-A^2)(x-y)^2$. 
For simplicity of the argument, we shall restrict ourselves to the $F_0>0$ case. 
In terms of these coordinates, the metric (\ref{AdS}) reduces to the standard coordinates of AdS as
\begin{align}
\label{AdS1}
\D s^2= -\left(k+g^2 R^2\right)\D T^2+\frac{\D R^2}{k+g^2 R^2}+R^2 
\D \Sigma_k^2(w, \varphi) \,.
\end{align}

In the case of  $g^2<A^2$ with $k=0, -1$ and $F_0>0$, the metric fails to be Lorentzian, 
which we shall not pursue any further. 
For $g^2<A^2$ with $k=1$ and $F_0>0$, we set $R\to -i  R$ and $T \to i T$
in (\ref{corrdAdS}), yielding the static AdS metric in the hyperbolic chart
\begin{align}
\label{}
\D s^2 =-\left(-1+g^2  R^2\right)\D T^2+\frac{\D R^2}{-1+g^2 R^2}+ R^2 \left(
\frac{\D w^2}{w^2-1}+(w^2-1)\D \varphi^2
\right) \,. 
\end{align}

For $A^2=g^2$, it turns out that only the $k\ne 0$ case provides the nondegenerate metric. Under this condition, we perform the following coordinate transformation
\begin{align}
\label{}
z=\frac{x-y}{y}\,, \qquad 
\rho =-\frac{\sqrt{1-kx^2}}{y} \,, 
\end{align}
yielding
\begin{align}
\label{AdS2}
\D s^2=\frac{1}{g^2 z^2}\Big(-k \D t^2+\D z^2+k^{-1}\D \rho^2+\rho^2 \D \varphi^2 \Big)\,. 
\end{align}
Since the metric in the parenthesis corresponds to the Minkowski spacetime, 
the spacetime (\ref{AdS2}) reduces to AdS written in the Poincar\`e coordinates.

\subsection{$A=0$ case: hairy black hole}
\label{sec:hairyBH}

Since the parameter $A$ measures the acceleration of a fiducial observer, 
let us next focus on the solution of vanishing acceleration. Setting $A=0$ in (\ref{sol1rx}), the solution reads
\begin{align}
\label{A0metric}
\D s^2=& \, 
-f(r)^{\frac{1-\alpha^2}{1+\alpha^2}}\Delta_r(r)\D\tau^2+\frac{\D r^2}
{f(r)^{\frac{1-\alpha^2}{1+\alpha^2}}\Delta_r(r)}
+r^2 f(r)^{\frac{2\alpha^2}{1+\alpha^2}}
\D \Sigma_k^2(x, \varphi) \,, \\
 \phi=&-\frac{2\alpha}{1+\alpha^2} \log f(r)\,, 
 \qquad A^0=\frac{q_0x}{\sqrt{1+\alpha^2}}\D \varphi\,, \qquad 
A^1 = \frac{q_1\alpha x}{\sqrt{1+\alpha^2}}\D \varphi\,, 
\end{align}
where 
\begin{align}
\label{}
\Delta_r (r)= & \, k-\frac{q_0^2}{r_0 r}+\frac{q_1^2}{r_0 r f(r)}+g^2 \left(r^2 -\frac{3\alpha^2-1}{1+\alpha^2} r_0 r +\frac{(\alpha^2-1)(3\alpha^2-1)}{(1+\alpha^2)^2}r_0^2\right)\,.
\end{align}
This two charged solution has been derived in \cite{Anabalon:2020pez}.\footnote{
Set $\alpha_{\rm there}=g$, $L_{\rm there}=1/g$, $x_{\rm there}=f(r)^{\frac{1-\alpha^2}{1+\alpha^2}}$
and $\eta_{\rm there}=(1+\alpha^2)/[gr_0(1-\alpha^2)]$ in eq. (3.25) of \cite{Anabalon:2020pez}. 
} 
As evident from the metric form, 
the locus of the event horizon is the largest root $r_+$ of $\Delta _r(r)=0$. If the event horizon 
conceals both of the curvature singularities at $r=0$ and $r=r_0$, the solution (\ref{A0metric}) is 
qualified as a static black hole in AdS.

To start with, 
it is instructive to see  the asymptotic behavior of the solution (\ref{A0metric}). 
In terms of the areal radius $S(r)=r f(r)^{\frac{\alpha^2}{1+\alpha^2}}$, 
the metric and the scalar field are expanded around $r\to \infty$ as 
\begin{align}
\label{}
\D s^2\simeq&\, -\left(k-\frac{2M}{S}+g^2S^2\right)\D \tau^2
+\frac{\D S^2}{k+\gamma -2M'/S+g^2 S^2}+S^2 \D \Sigma_k^2(x, \varphi)\,, \notag \\
\phi\simeq &\,\frac{\phi_1}{S}+\frac{\phi_2}{S^2}\,,
\end{align}
where 
\begin{align}
\label{mass0}
M=&\, \frac{(1-\alpha^2)r_0}{6(1+\alpha^2)^3} \left[3k(1+\alpha^2)^2+g^2r_0^2(3\alpha^2-1)(\alpha^2-3)\right]+\frac{q_0^2-q_1^2}{2r_0}\,, \\
M'=&\, M+\frac{2g^2r_0^3\alpha^2(\alpha^2-1)}{3(1+\alpha^2)^3}\,, \\
\gamma=&\, \frac{\alpha^2g^2r_0^2}{(1+\alpha^2)^2 } \,, 
\end{align}
and 
\begin{align}
\label{}
\phi_1=\frac{2r_0\alpha}{1+\alpha^2}\,, \qquad \phi_2=-\frac{r_0^2 \alpha(\alpha^2-1)}{(1+\alpha^2)^2} \,.
\end{align}
The unfamiliar term $\gamma$ originates from the existence of slower fall-off mode $\phi_1$ of the scalar field
around the AdS vacuum \cite{Hertog:2004dr}, which is intricately 
related to the notion of multi-trace deformations of conformal field theory.
According to the prescription given in \cite{Hertog:2004dr,Papadimitriou:2007sj}, 
the physical mass is given by $M$, rather than $M'$. This outcome is convincingly justifiable by the 
first law of black hole thermodynamics
\begin{align}
\label{}
\delta M=\frac{\kappa}{8\pi}\delta {\rm Area}+\Phi_0\delta Q_0+\Phi_1 \delta Q_1 \,, 
\end{align}
where $\kappa$ and ${\rm Area}$ correspond, respectively, to the surface gravity--associated with the 
time translation $\partial/\partial \tau$--and the area of the 
event horizon $r=r_+$
\begin{align}
\label{}
\kappa=\frac 12 f(r_+)^{\frac{1-\alpha^2}{1+\alpha^2}}\Delta_r'(r_+)\,, \qquad 
{\rm Area}=4\pi S^2(r_+) \,.
\end{align}
Here we have assumed $\Sigma_k$ to be compact with area $4\pi$. 
The magnetic charges and magnetostatic potentials are given by 
\begin{align}
\label{}
Q_0=\frac{q_0}{\sqrt{1+\alpha^2}}\,, \qquad 
Q_1=\frac{\alpha q_1}{\sqrt{1+\alpha^2}}\,, \qquad 
\Phi_0=\frac{Q_0}{r_+} \,, \qquad 
\Phi_1=\frac{Q_1}{r_+f(r_+)}\,. 
\end{align}
Thermodynamic aspects of this solutions have been discussed in \cite{Anabalon:2021smx}. 

Since $\Delta_r(r)$ allows at most four real roots, 
the classification of horizons requires a formidable work, which we shall not pursue further 
in this paper. In lieu of this, 
let us consider a simpler case in which the solution is neutral $q_0=q_1=0$. 
In this case, the solution has been already constructed in \cite{Faedo:2015jqa} and rederived via Mcvittie's ansatz in \cite{Nozawa:2020gzz}. 
The solution (\ref{A0metric}) asymptotically $r\to \infty$ approaches to AdS and admits a parameter range in which
the event horizon $r=r_+$ exists at the largest root $r_+$ of $\Delta(r)=0$ outside the curvature singularities at 
$r=0$ and $r=r_0$. 
We summarize the results of \cite{Faedo:2015jqa} in table~\ref{table:BHrange}.\footnote{
Since the scalar field contributes nontrivially to the gravitational Hamiltonian, 
the positivity of the mass $M$ is far from clear to date, when the scalar field
displays the slow fall-off at infinity. Nevertheless, 
it deserves to remark that the mass $M$ given in (\ref{mass0}) is indeed positive
for the solution (\ref{A0metric}), whenever the horizon exists for $k=1$. 
} 
It therefore follows that the neutral solution describes a 
hairy black hole in AdS. This implies the violation of the uniqueness conjecture of static black holes, 
since the present theory (\ref{Lag}) admits a Schwarzschild-AdS black hole without a scalar hair.

\begin{table}
  \centering 
\begin{tabular}{c||c|c|c|c|c|c|c}
\hline
& $0<\alpha<1/{\sqrt 3}$ & $\alpha=1/{\sqrt 3}$ & $1/{\sqrt 3}<\alpha<1$&
$\alpha=1$& $1<\alpha<{\sqrt 3}$ & $\alpha=\sqrt 3$& $\alpha>{\sqrt 3}$
\\ \hline\hline
  $k=1$ & -- & -- & $gr_0<-f_1(\alpha)$& -- &
  $gr_0>f_2(\alpha)$&-- & --  \\ \hline
  $k=0$ & -- &-- & $r_0<0 $ & --&$r_0>0$ &-- & -- \\ \hline
$k=-1$   & $-2f_3(\alpha)\le gr_0<f_2(\alpha)$ &\multicolumn{2}{c|}{$gr_0<f_2(\alpha)$} & $r_0\in \mathbb R$ 
&\multicolumn{2}{c|}{$gr_0>-f_1(\alpha)$} & 
$-f_1(\alpha)<gr_0 \le 2f_3(\alpha)$
 \\
\hline
\end{tabular}
\caption{The parameter range under which the solution (\ref{A0metric}) describes a black hole 
which is regular on and outside the event horizon $r_+>{\rm max}(0, r_0)$. $f_{1-3}(\alpha)$ are defined by
$f_1(\alpha)=(1+\alpha^2)/{\sqrt{|(\alpha^2-1)(1-3\alpha^2)|}}$, 
$f_2(\alpha)=(1+\alpha^2)/{\sqrt{|(\alpha^2-1)(3-\alpha^2)|}}$, 
$f_3(\alpha)=(1+\alpha^2)/{\sqrt{|(\alpha^2-3)(3\alpha^2-1)|}}$. 
Thanks to the symmetry (\ref{sym2}), we have tabulated only the range  $\alpha>0$. 
}
\label{table:BHrange}
\end{table}

\subsection{$g=0$ case}

Due to $V\propto g^2$, setting $g=0$ gives rise to the massless scalar. 
In this case, one obtains  the dilatonic C-metric in the asymptotically flat space in a broad sense.  
Since the term of non-integer power of $h(x)$ drops out of $\Delta_x$, 
the resulting metric is akin to the singly charged dilatonic C-metric in  \cite{Dowker:1993bt}.

\subsection{Double Wick rotation and comparison with the literature}
\label{sec:flip}

Let us consider the general structure functions (\ref{general}). 
We then perform the following double Wick rotation
\begin{align}
\label{flip}
x=\hat y\,, \qquad  y=\hat x \,, 
\qquad 
t= i \hat \varphi \,, \qquad  \varphi =i\hat t \,, \qquad 
q_0=-i \hat q_0\,, \qquad q_1=-i \hat q_1 \,. 
\end{align}
This amounts to the simultaneous interchange of the role of ($x, y$) and ($t, \varphi$). 
Note that the flip of $t$ and $ \varphi$ involves the double Wick rotation to keep the Lorentzian signature.
Then, the metric reduces to 
another family of C-metrics 
\begin{align}
\label{}
\D s^2=&\,\frac 1{A^2(\hat x-\hat y)^2} \left[
h(\hat x)^{\frac{2\alpha^2}{1+\alpha^2}}\left(-h(\hat y)^{\frac{1-\alpha^2}{1+\alpha^2}}\hat \Delta_{\hat y}(\hat y)\D \hat t^2+\frac{\D\hat  y^2}
{h(\hat y)^{\frac{1-\alpha^2}{1+\alpha^2}}\hat \Delta_{\hat y}(\hat y)}\right)\right.
\notag \\ & \left.+h(\hat y)^{\frac{2\alpha^2}{1+\alpha^2}}
\left(\frac{\D \hat x^2}{h(\hat x)^{\frac{1-\alpha^2}{1+\alpha^2}}\hat \Delta_{\hat x}(\hat x)}
+h(\hat x)^{\frac{1-\alpha^2}{1+\alpha^2}}\hat \Delta_{\hat x}(\hat x) \D \hat \varphi^2 
\right)
\right]\,, \label{sol2}\\
 \phi=&\frac{2\alpha}{1+\alpha^2} \log \left(\frac{h(\hat y)}{h(\hat x)}\right)\,, 
 \qquad A^0=\frac{\hat q_0 \hat y}{\sqrt{1+\alpha^2}}\D \hat t \,, \qquad 
 A^1=\frac{\alpha \hat q_1 \hat y}{\sqrt{1+\alpha^2}h(\hat y)}\D \hat t\,, 
\end{align}
where 
the structure functions are 
written explicitly as
\begin{align}
\label{hatDelta}
\hat \Delta_{\hat y}(\hat y) \equiv \,& a_0+2a_1 \hat y+a_2 \hat y^2
+\frac{A\hat q^2_0}{r_0}\hat y^3-\frac{A\hat q_1^2\hat y}{r_0 h(\hat y)} +\frac{g^2}{A^2} h(\hat y)^{\frac{3\alpha^2-1}{1+\alpha^2}}\,, \\
\hat \Delta_{\hat x} (\hat x)\equiv \,&  -a_0 -2 a_1 \hat x-a_2 \hat x^2-\frac{A\hat q_0^2}{r_0}\hat x^3+\frac{A\hat q_1^2 \hat x}{r_0 h(\hat x)}\,, 
\end{align}
As is obvious from the construction, this solution solves the field equations for the system (\ref{Lag}) as well. 
The differences between the metrics (\ref{sol1}) and (\ref{sol2}) are the precise form of the structure 
functions, the sign of the scalar field and the electric/magnetic configurations of the gauge potentials. 
As it turns out, 
the solution  (\ref{sol2}) incorporates the one found in \cite{Lu:2014ida,Lu:2014sza}. 
To facilitate the comparison with the notation in~\cite{Lu:2014ida}, 
it is opportune to introduce
\begin{align}
\label{}
\hat r=-\frac 1{A\hat y}\,, \qquad 
\hat \tau =\frac{\hat t}{A} \,, \qquad 
\hat \Delta_{\hat r} (\hat r)=\, 
a_2-2a_1 A \hat r+a_0 A^2 \hat r^2-\frac{\hat q^2_0}{r_0 \hat r}
+\frac{\hat q_1^2 }{r_0r f(\hat r)}
+g^2 \hat r^2 f(\hat r)^{\frac{3\alpha^2-1}{1+\alpha^2}}\,. 
\end{align}
Then, the metric (\ref{sol2}) reduces to the `hatted' form of (\ref{sol1rx}), for which the $A\to 0$ limit can be taken.  
Using the freedom corresponding to (\ref{shift}), 
one can fix the parameters as 
\begin{align}
\label{para:sol2}
a_0=-1 \,, \qquad a_1=0 \,, \qquad a_2 =k \,. 
\end{align}
Thus, one ends up with 
\begin{align}
\label{}
\hat \Delta_{\hat r} (\hat r)=k -A^2 \hat r^2 -\frac{\hat q^2_0}{r_0 \hat r}
+\frac{\hat q_1^2 }{r_0\hat r f(\hat r)}
+g^2 \hat r^2 f(\hat r)^{\frac{3\alpha^2-1}{1+\alpha^2}}\,, \qquad 
\hat \Delta_{\hat x}(\hat x)=1-k\hat x^2-\frac{A\hat q^2_0}{r_0}\hat x^3 
+\frac{A \hat q_1^2 \hat x^2}{r_0 h(\hat x)}
\,. 
\end{align} 
We see that the solution (\ref{sol2}) with $k=1$ reduces, after some shift and scaling transformations (\ref{shift}),  
to  the one found in \cite{Lu:2014ida,Lu:2014sza}. It is worthy of remark that the choice of parameters 
$a_{0,1,2}$ required by the existence of the $A\to 0$ limit (\ref{para:sol2}) differs from the previous one (\ref{para:sol1}).

Setting $r_0=\hat q_{0,1}=0$, the metric (\ref{sol2}) with (\ref{para:sol2}) reduces to AdS, while the solution in the $A\to 0$ limit leads to 
\cite{Anabalon:2012ta,Feng:2013tza}
\begin{align}
\label{A0metric2}
\D s^2=& \, 
-f(\hat r)^{\frac{1-\alpha^2}{1+\alpha^2}} \hat \Delta_{\hat r}(\hat r)\D \hat \tau^2+\frac{\D \hat r^2}
{f(\hat r)^{\frac{1-\alpha^2}{1+\alpha^2}} \hat \Delta_{\hat r}(\hat r)}
+\hat r^2 f(\hat r)^{\frac{2\alpha^2}{1+\alpha^2}}
\D\Sigma_k^2 (\hat x, \hat \varphi) \,, \\
\phi=&+\frac{2\alpha}{1+\alpha^2} \log f(\hat r)\,, \qquad 
A^0=-\frac{\hat q_0 }{\sqrt{1+\alpha^2}\hat r}\D \hat \tau\,, \qquad 
A^1=-\frac{\alpha \hat q_1 }{\sqrt{1+\alpha^2}\hat r f(\hat r)}\D \hat \tau\,, 
\end{align}
where
\begin{align}
\label{}
 \hat \Delta_{\hat r}(\hat r)=k-\frac{\hat q^2_0}{r_0 \hat r}
+\frac{\hat q_1^2 }{r_0\hat r f(\hat r)}+g^2 \hat r^2f(\hat r)^{\frac{3\alpha^2-1}{1+\alpha^2}}\,.
\end{align} 
By the parallel argument laid out in previous subsections, $r_0$ corresponds to the mass parameter and 
$k$ governs the topology of the horizon. The solution (\ref{A0metric2}) asymptotically tends to AdS as $\hat{r}\to \infty$ with the mass
\begin{align}
\label{}
M=\frac{r_0 k (1-\alpha^2)}{2(1+\alpha^2)}+\frac{\hat q_0^2-\hat q_1^2}{2r_0}\,. 
\end{align}

It is obvious to see that the neutral solution (\ref{A0metric2}) does not allow an event horizon 
of a black hole for $k\ge 0 $, by virtue of $\hat \Delta_{\hat r}(\hat r)=k+g^2 \hat r^2f(\hat r)^{\frac{3\alpha^2-1}{1+\alpha^2}}>0$. 
This is in sharp contrast 
to the solution (\ref{A0metric}) (see table~\ref{table:BHrange}). It follows that the solutions (\ref{A0metric}) and (\ref{A0metric2}) describe genuinely distinct 
family of physical spacetimes, even though they are related by (\ref{flip}).
This is reminiscent of the fact \cite{Kunduri:2007vf} that the Wick rotation of the near-horizon geometry of a dipole black ring 
\cite{Emparan:2004wy} gives rise to the Kaluza-Klein black hole \cite{Ishihara:2005dp} in five dimensional Einstein-Maxwell theory.

Since the discovery of the solution (\ref{A0metric}), it has remained open why the 
same theory admits two discrete family of static solutions (\ref{A0metric}) and (\ref{A0metric2}) of black hole type. 
These two solutions are not related to the ordinary electromagnetic duality.\footnote{This is because 
the metrics (\ref{A0metric}) and (\ref{A0metric2}) are solutions for the {\it same} Lagrangian (\ref{Lag}).
The electromagnetic duality $(F^0, F^1)\to (\star F^0{}', \star  F^1{}')=(e^{-\alpha \phi}F^0, e^{\frac 1\alpha \phi}F^1)$ is not a symmetry for the Lagrangian but for equations of motion of gauge fields.  If the potential of the scalar field vanishes, the transformation
$(F^0, F^1)\to (\star F^0{}', \star  F^1{}')=(e^{-\alpha \phi}F^0, e^{\frac 1\alpha \phi}F^1)$ can be compensated by the sign flip of the 
scalar field and the new solution falls into the same theory. Obviously, this is not the case since  the potential is not an even function.
Thus, although an electrically charged solution is obtained by performing the electromagnetic duality to the solution (\ref{A0metric}), it is a solution to a different theory. In ref. \cite{Anabalon:2017yhv}, the authors clarified the existence of two solutions in terms of the generalized electromagnetic 
duality, at the price of introducing 
magnetic gaugings $g_{\mathbb M}=(g_I, g^I)$ and $F^{\mathbb M}=(F^I, F_I)$. This formulation restores the symplectic covariance and 
is related to the $\omega$-deformation of ${\cal N}=8$ gauged supergravity.
}
We have clarified above that these seemingly different solutions are indeed related by (\ref{flip}), which cannot be captured 
unless one introduces the acceleration parameter $A$.

\subsection{Hairless case}

Since the origin of the potential is an extremum, one can truncate the 
theory to $\phi=0$ provided $F^I=0$. 
The potential reads $V=-3g^2$, corresponding to the negative cosmological constant
with AdS radius $g^{-1}$. It follows that the present theory (\ref{Lag}) also 
possesses the ordinary hairless C-metric
\begin{align}
\label{sol3}
\D s^2=\frac 1{A^2 (x-y)^2}\left(
-\check \Delta_y(y)\D t^2+\frac{\D y^2}{\check\Delta_y(y)}
+\frac{\D x^2}{\check\Delta_x(x)}+\check\Delta_x(x)\D \varphi^2
\right) \,,
\end{align}
where
\begin{align}
\label{vac:Deltayx}
\check \Delta_y(y)= \frac{g^2}{A^2}-1+k y^2+2Am y^3 \,, \qquad 
\check\Delta_x(x)=1-k x^2 -2 A m x^3 \,. 
\end{align}
Here $k=0, \pm 1$. 
The causal structure of this solution with 
$k=1$ and $m>0$ has been discussed by
\cite{Dias:2002mi,Podolsky:2002nk}. In this $\Lambda$-vacuum case, both of 
the structure functions $\check \Delta_y$ and $\check \Delta_x$ are 
cubic in each variable. This means that the double Wick rotation  (\ref{flip}) is trivial.

It is worthy of mentioning that the hairless solution (\ref{sol3}) is {\it not} derived from 
(\ref{sol1}) or (\ref{sol2}), since the only scheme to set the scalar field to be constant 
for the latter two solutions is $r_0=0$, eventuating in AdS. 
This implies that a more general solution encompassing all of these three
distinct solutions should exist. 
Actually, families of numerical solutions were found in the double well potential case, and
it was reported that such solutions exist generally around the top of the upwardly convex potential \cite{Torii:2001pg}.
The relation between these solutions and our hairy black hole solutions are still unclear, but
it may be possible to understand the black hole solutions in a unified manner.
We defer the extensive search of the solution of this sort 
to a future work.

\section{Physical properties of the solution}
\label{sec:property}

The discussion in the previous section shows that the present theory (\ref{Lag}) with $F^I=0$ enjoys
three family of C-metrics (\ref{sol1}), (\ref{sol2}) and (\ref{sol3}). It follows that the rich variety of black hole solutions 
in this theory is not only limited to ordinary static black holes, but also persists to the accelerating solutions. 
This encourages further motivation for investigating their physical properties.

As demonstrated, our C-metric (\ref{sol1}) recovers the hairy black hole, while the solution (\ref{sol2})
does not have horizons for $q_{0,1}=0$ and $k=1$. It is then reasonable to infer that the solution (\ref{sol1}) 
describes a C-metric supported solely by a scalar hair. 
To demonstrate this prospect rigorously, we need to clarify the global causal structure. 
This is the prime purpose of the current section. 

\subsection{Conical singularity}

We are interested in the case where the metric has a Lorentzian signature ($-, +, +, +$). 
This restricts the domain of $x$ to the range $\Delta_x(x) \ge 0 $. The precise form of 
$\Delta_x(x) $ is sensitive to the value of all seven parameters $g, \alpha $ and $r_0, A, q_{0,1}, k$. 
We focus on the case in which $\Delta_x(x)$ admits at least two real roots $x_\pm$ 
\begin{align}
\label{xrange}
\Delta_x (x_\pm )=0 \, , \qquad 
\Delta_x (x)>0 ~~ (x_- < x < x_+) \, . 
\end{align}
In this case, the two dimensional surface spanned by ($x, \varphi$) becomes compact. 
For instance, the special case of $\alpha =1$, $\sqrt 3$, $1/\sqrt 3$ gives 
$\Delta _x(x)=1-k x^2$ for $q_{0,1}=0$, requiring  $k=1$.

Possible conical singularities at $x=x_-$ 
can be avoided, provided $\varphi_- =\frac 12 h(x_-)^{\frac{1-\alpha^2}{1+\alpha^2}}|\Delta_x' (x_-)|\varphi$ 
has a canonical $2\pi$ period. 
One can determine the periodicity of $\varphi$ at $x=x_+$ in an analogous fashion, and it turns out to be  
$2\pi -\delta$, where
\begin{align}
\label{}
\delta =2 \pi \left(1-\frac{h(x_+)^{\frac{1-\alpha^2}{1+\alpha^2}}|\Delta_x'(x_+)|}{h(x_-)^{\frac{1-\alpha^2}{1+\alpha^2}}
|\Delta_x'(x_-)|}\right)\,.
\end{align}
Generically, there emerges a conical singularity $\delta \ne 0$ at $x=x_+$. 
The positive $\delta$ corresponds to the conical deficit, while the negative
$\delta$ corresponds to the conical excess.

An exceptional case is $\alpha =1$ with $q_{0,1}=0$, for which 
the two dimensional surface 
$\D s_2^2=\D x^2/(h(x)^{\frac{1-\alpha^2}{1+\alpha^2}}\Delta _x(x))
+h(x)^{\frac{1-\alpha^2}{1+\alpha^2}}\Delta _x(x)\D \varphi^2$ with $k=1$ 
is simplified to the standard metric of the unit two sphere
$\D s_2^2=\D \theta^2+\sin^2\theta \D \varphi^2$, where 
$x=\cos\theta$. It follows that the conical singularity at the north and south poles of 
$S^2$ can be completely cured. It is the background scalar field that provides the acceleration. 
However, it turns out that this case  
describes a naked singularity. 

\subsection{Infinity and singularity}

The coordinate $x$ might be regarded as a directional cosine ($x=\cos\theta$), but the  coordinate $r=-1/(Ay)$ may not be taken too literally as an ordinary radial coordinate. To see this more concretely, let us consider  `radial' null geodesics obeying 
\begin{align}
\label{}
-h(y)^{\frac{1-\alpha^2}{1+\alpha^2}}\Delta_y(y)\dot t^2+\frac 1{h(y)^{\frac{1-\alpha^2}{1+\alpha^2}}\Delta_y(y)}\dot y^2 =0 \,, \qquad E=\frac{h(x)^{\frac{2\alpha^2}{1+\alpha^2}}h(y)^{\frac{1-\alpha^2}{1+\alpha^2}}\Delta_y (y)}{A^2(x-y)^2} \dot t\,, 
\end{align}
where the dot denotes the derivative with respect to the affine parameter $\lambda$  and 
$E$ is a constant corresponding to the energy of null rays, respectively. Upon integration, we have
\begin{align}
\label{}
\pm EA^2 (\lambda-\lambda _0) = \int \frac{h(x)^{\frac{2\alpha^2}{1+\alpha^2}}}{(x-y)^2}\D y = \frac{h(x)^{\frac{2\alpha^2}{1+\alpha^2}}}{x-y} \,. 
\end{align} 
It follows that $r=+\infty $ ($y=0$) can be reached within a finite affine time for these null geodesics. 
One also recognizes that the surface $x=y$ corresponds to infinity. This enforces us to work with
coordinate $y$, instead of $r$, to reveal the global causal structure.

The spacetime singularities are characterized by the blow-up of 
curvature invariants. 
The scalar curvature and the Kretschmann invariant diverge at 
\begin{align}
\label{}
y=\pm \infty\,, \qquad y=-\frac 1{Ar_0} \,, \qquad 
x=-\frac 1{Ar_0}\,. 
\end{align}
Since we are now paying attention only to 
the finite range of $x$, other plausible singularities $x=\pm \infty$ are not our concern here. 
A minimal requirement for regularity of the solution is that $x=-1/(Ar_0)$ lies outside the range of $x_-\le x\le x_+$ and 
the Killing horizons cover singularities $y=\pm \infty$ and $y=-1/(Ar_0)$.

To see the structure of $y={\rm const.}$ surface such as singularities and infinity for fixed $x$, 
it suffices to focus on the two dimensional portion of the spacetime 
\begin{align}
\label{}
\D s^2_2=-\Delta_y(y)\D t^2+\frac{\D y^2}{\Delta_y(y)}=-\Delta_y(y(y_*))\left(\D t^2-\D y_*^2\right)\,, 
\end{align}
where $y_*=\int \Delta_y^{-1}\D y $ is the analogue of the tortoise coordinate. 
Since this metric is manifestly conformal to the two dimensional Minkowski metric, 
one can immediately extract   the causal structure of the $y={\rm const.} $ surface. 
If $y_*$ diverges, the corresponding surface is null. 
If $y_*$ is finite, the corresponding surface is spacelike (timelike)
for $\Delta_y(y(y_*)) <0~ (>0)$.

\begin{figure}[t]
\begin{center}
\includegraphics[width=10cm]{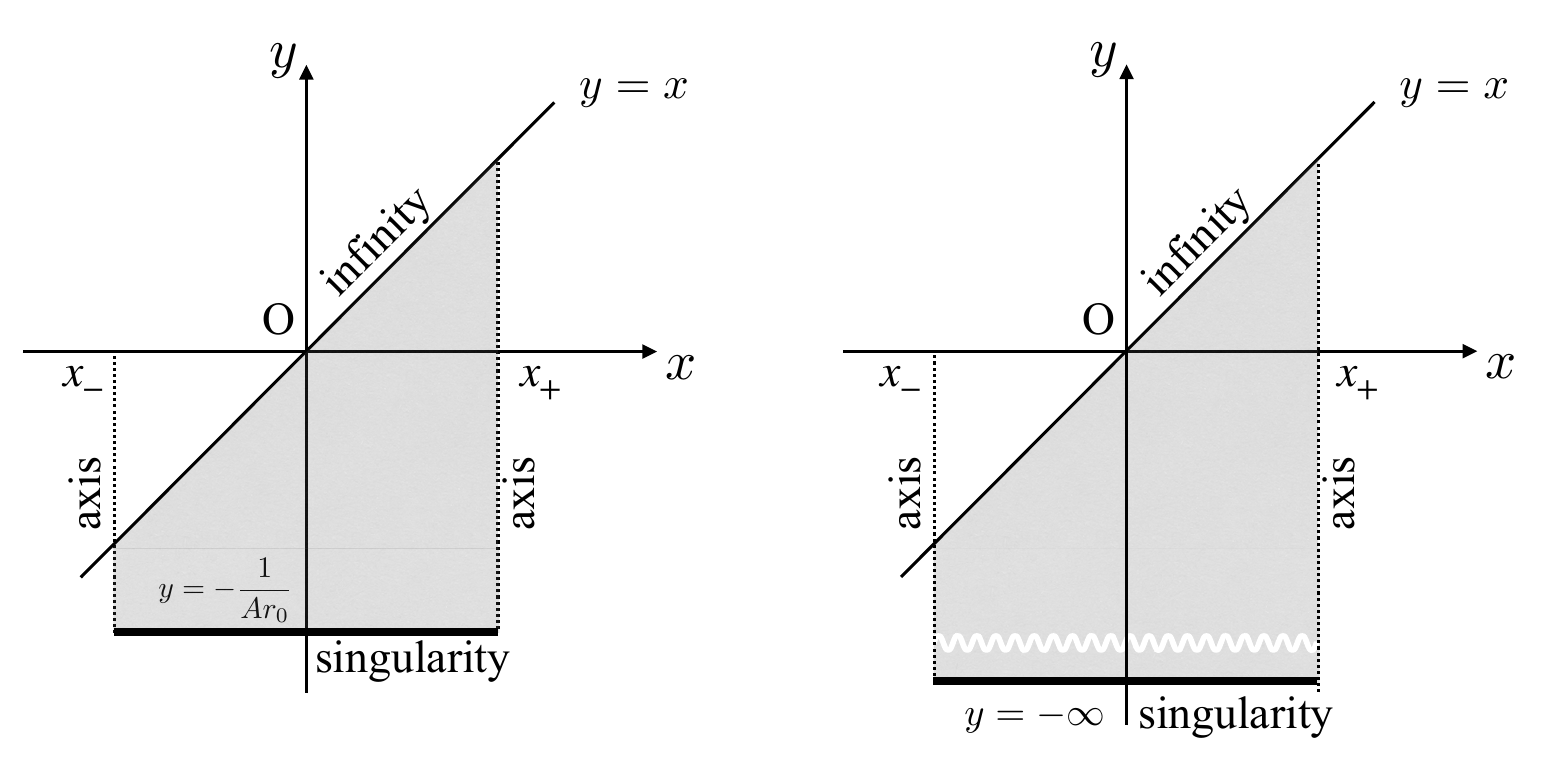}
\caption{Coordinate region for $r_0>0$ (left) and $r_0<0$ (right). }
\label{fig:coord}
\end{center}
\end{figure}

Since $x-y=0$ corresponds to the asymptotic infinity, 
we postulate that the physical region is enclosed by
\begin{align}
\label{}
x-y\ge 0 \,, \qquad 
h(x)> 0\,, \qquad h(y)>0 \,, \qquad 
\Delta_x(x)\ge 0 \,. 
\end{align}
We shall not consider the 
$x-y<0$ region, since it is simply achieved by the simultaneous sign flip of 
$x, y$ and $r_0$. 
The coordinate domain under consideration is visualized in figure~\ref{fig:coord}.

\subsection{Causal structure:  neutral case}

Let us first explore the causal structure of the neutral case $q_{0,1}=0$ with $k=1$.
The motivation comes from the fact that the solution (\ref{sol1})
gives rise to the hairy black hole in the limit $A\to 0$ (see section \ref{sec:hairyBH}),
in contrast to the case (\ref{sol2}). It is then tempting to envisage that the 
$q_{0,1}=0$ solution (\ref{sol1}) would describe the hairy C-metric in AdS.

The Killing horizons appear when 
the Killing vector for the time translation $\partial/\partial t$ becomes null. 
This occurs at $\Delta_y=0$, which admits at most two distinct roots which we denote
$y_-<y_+$, since $\Delta_y$ is quadratic in $y$ (\ref{Deltays}). 
In the hereafter, we focus on the case where these horizons exist and are 
nondegenerate.

We now proceed to discuss the global structure. 
For this purpose, it is helpful to observe the following relation
\begin{align}
\label{Deltaxy}
\Delta _x(x)=\frac{g^2}{A^2}h(x)^{\frac{3\alpha^2-1}{1+\alpha^2}}-\Delta_y(x) \,. 
\end{align}
This means that the intersecting points of the functions $(g^2/A^2)h(x)^{\frac{3\alpha^2-1}{1+\alpha^2}}$ 
and $\Delta_y(x) $ correspond to the axes $x=x_\pm$. 
By the requirement of Lorentzian signature $\Delta_x(x)\ge 0$, the permitted region is 
 $(g^2/A^2)h(x)^{\frac{3\alpha^2-1}{1+\alpha^2}}\ge \Delta_y(x)$. 
If $\Delta_y(x)$ is convex upward, the Lorentzian signature is assured 
for $x<x_-$ or $x>x_+$. This makes the surface spanned by $x-\phi$ coordinates 
noncompact, which we are not concerned with. It follows that $\Delta_y(x)$ must be convex downward.

\begin{figure}[t]
\begin{center}
\includegraphics[width=12cm]{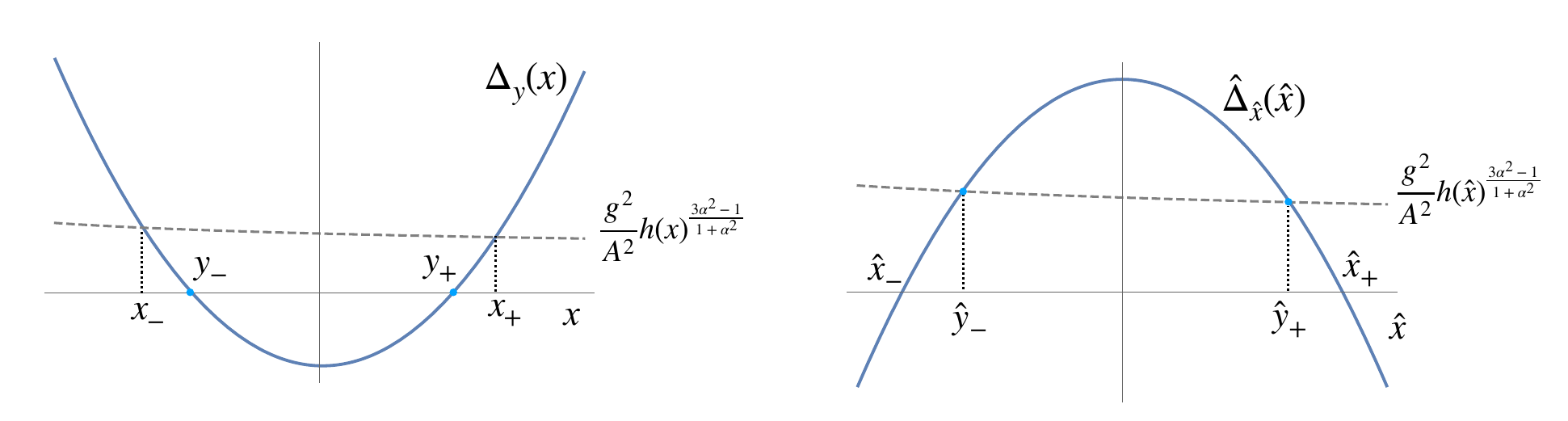}
\caption{Plots of $\Delta_y(x)$ and $(g^2/A^2)h(x)^{\frac{3\alpha^2-1}{1+\alpha^2}}$ (left) 
and $\hat \Delta_{\hat x}(\hat x)$ and $(g^2/A^2)h(\hat x)^{\frac{3\alpha^2-1}{1+\alpha^2}}$ (right). 
The parameters are chosen such that $g=k=1$, $A=3/2$, $r_0=1/4$, $\alpha=1/2$ and 
$q_{0,1}=\hat q_{0,1}=0$. In either case, $x_-<y_-<y_+ < x_+$. }
\label{fig:Deltay}
\end{center}
\end{figure}

Typical behaviors of functions $\Delta_y(x)$ and $(g^2/A^2)h(x)^{\frac{3\alpha^2-1}{1+\alpha^2}}$
are plotted in the left of figure \ref{fig:Deltay}. These functions may have three intersections, but this is not essential 
for the present discussion. 
To extract a useful relation, we remark the subsequent features:
\begin{itemize}
  \item $(g^2/A^2)h(x)^{\frac{3\alpha^2-1}{1+\alpha^2}}$ is 
positive-semidefinite and monotonically increasing (decreasing) for $(3\alpha^2-1)r_0 >0~ (<0)$. 
  \item $\Delta_y(x)$ is convex downward with $\Delta_y(y_\pm)=0$.
  \item $x$ takes values in the bounded domain $x_-\le x\le x_+$ satisfying 
  $\Delta_x(x)=(g^2/A^2)h(x)^{\frac{3\alpha^2-1}{1+\alpha^2}}-\Delta_y(x)\ge0$.
  \item  Conditions $-1/(Ar_0)<x_-$ for $r_0>0$ and $x_+<1/(A|r_0|)$ for $r_0<0$ must be 
  satisfied to evade the naked singularity. 
\end{itemize}
Inspecting these aspects, we arrive at the universal relation
\begin{align}
\label{ypmxpm}
x_-<y_-<y_+<x_+ \,, 
\end{align}
regardless of the parameters.

\begin{figure}[t]
\begin{center}
\includegraphics[width=13cm]{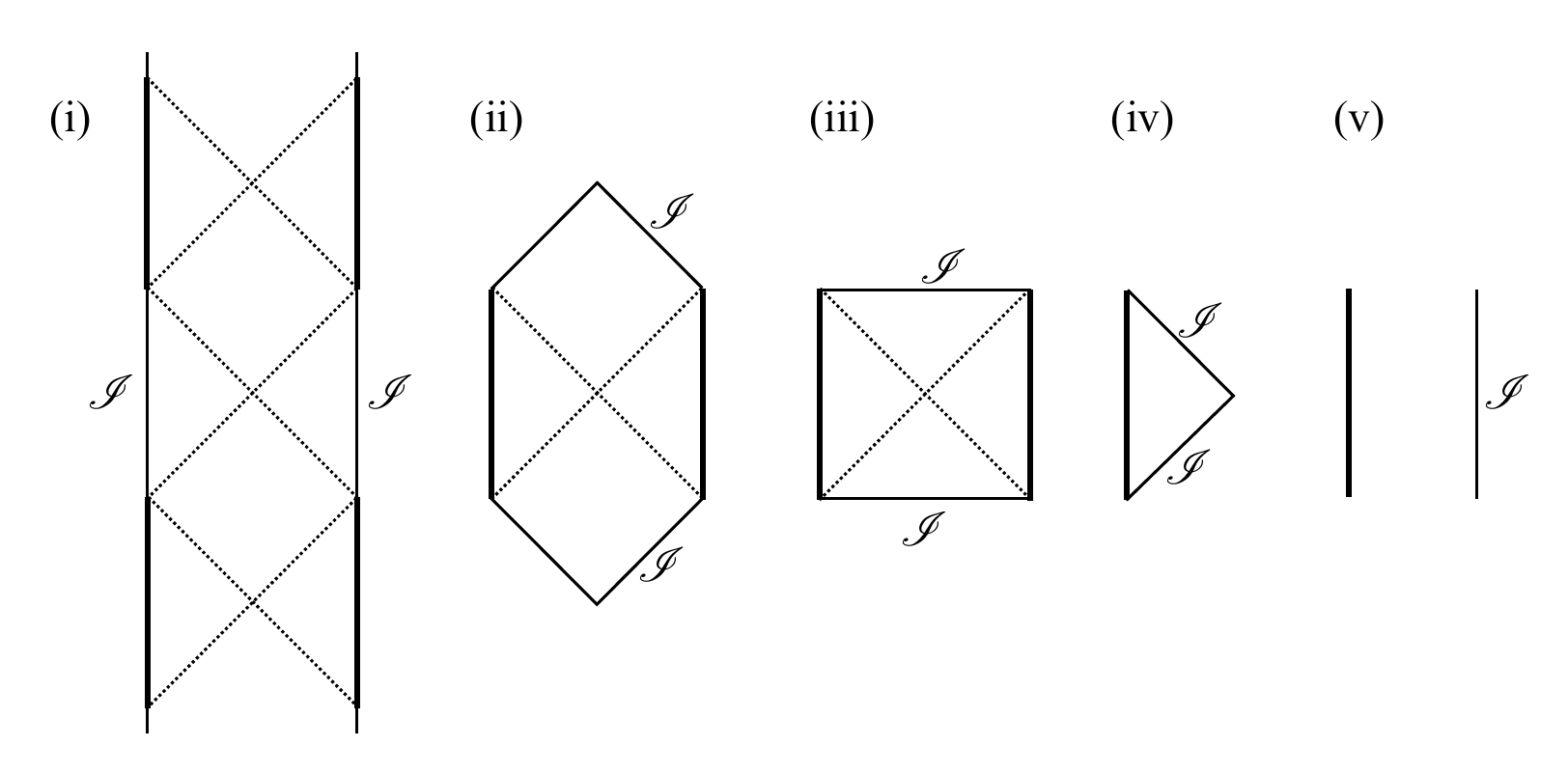}
\caption{Penrose diagrams of the $k=1$ neutral C-metric.
The thick lines denote the curvature singularities. 
}
\label{fig:PD1}
\end{center}
\end{figure}

Recalling that the allowed coordinate region is 
$x>y$ with $x=y$ being conformal infinity, 
the Penrose diagram is designed as follows: 
\begin{itemize}
  \item[(i)] $y_+<x\le x_+$. In this case, we have two nondegenerate Killing horizons $y=y_\pm$ and infinity is timelike. 
  The Penrose diagram is the same as the Reissner-Nordstr\"om-AdS black hole [fig \ref{fig:PD1}-(i)]. 
  \item[(ii)] $x=y_+$. We have a single horizon at 
  $y=y_-$ and infinity corresponds to the null surface [fig \ref{fig:PD1}-(ii)].
  \item[(iii)] $y_-<x<y_+$. We have a single horizon at 
  $y=y_-$ and infinity becomes spacelike [fig \ref{fig:PD1}-(iii)].
    \item[(iv)] $x=y_-$. We do not have any horizons and infinity is replaced by a null surface  [fig \ref{fig:PD1}-(iv)]. 
    \item[(v)] $x_-\le x<y_-$. No horizons are present and infinity alters to timelike [fig \ref{fig:PD1}-(v)]. 
\end{itemize}
Here, the singularity refers to either $y=-1/(A r_0)$ ($r_0>0$)
or $y=-\infty$ ($r_0<0$).

It turns out that the solution (\ref{sol1}) with $k=1$ and $q_{0,1}=0$ fails to describe 
the accelerating black holes in AdS. Rather, it corresponds to the accelerating naked singularity. 
The singularity is covered only from a part of angular directions $y_+<x\le x_+$ [case (i)], otherwise
the singularity is globally visible from future infinity.
This is a bit puzzling and unanticipated result, since 
the $A\to 0$ limit gives rise to a hairy black hole for some range of parameters, as shown in table~\ref{table:BHrange}. 
This is because the $A \to 0$ limit of the solution is discontinuous. 
In the $A\ne 0$ case, a stringent restriction is placed upon the causal structure because of 
the relation (\ref{ypmxpm}). In contrast, the $r$ and $x$ coordinates for the $A\to 0$ metric (\ref{A0metric}) are  free from this condition. 
As long as $\Delta_y(y)$ is a quadratic function of $y$, the constraint (\ref{ypmxpm}) is inevitable. 
The $\Lambda$-vacuum case, on the other hand, can circumvent this problem since $\check \Delta_y(y)$ is cubic [see (\ref{vac:Deltayx})]. 

The neutral limit of the solution (\ref{sol2}) with $k=1$
does not alleviate this problem. 
Considering that the functional relationship 
$\hat \Delta_{\hat x}(\hat x)=(g^2/A^2)h(\hat x)^{\frac{3\alpha^2-1}{1+\alpha^2}}-\hat \Delta_{\hat y}(\hat x)$
remains untouched and 
$\hat \Delta_{\hat x}(\hat x)$ is  convex upward with 
$\hat \Delta_{\hat x}(\pm 1)=0$, we have 
$\hat x_-=-1<\hat y_-<\hat y_+<1=\hat x_+$ (right picture in figure \ref{fig:Deltay}). 
This is the relation identical to (\ref{ypmxpm}), 
leading to the same Penrose diagram (figure \ref{fig:PD1}) as above. 
It follows that none of the neutral solutions (\ref{sol1}) and (\ref{sol2}) 
correspond to the hairy C-metrics.

\subsection{Causal structure: charged solution}

As revealed in the precedent subsection, 
the scalar field is not capable of supporting the regular configuration (modulo the conical singularity) of the C-metric.
We therefore examine the effect of charging up, for which $\Delta _y(y)=0$ admits at most four real roots. 
However, the exhaustive classification of the horizons and axes becomes intractable or, at the very best,  considerably cumbersome, 
in light of the fact that the solution involves seven parameters and $\Delta_x(x)$ contains a non-integer power function of $x$. 
For the discussion to be reasonably focused, we confine to the case 
$\alpha=1/\sqrt 3$ with $k=1$ and $q_1=0$, in which $\Delta_x(x)$ and $\Delta_y(y)$ are simplified to the cubic functions as
$\Delta_x(x)=1-x^2-(Aq_0^2/r_0)x^3$ and $\Delta_y(y)=g^2/A^2-\Delta_x(y)$. 
This  case is simple but nonetheless captures the essential feature of the global structure for other value of parameters.

Since the intractable term $h(x)^{\frac{3\alpha^2-1}{1+\alpha^2}}$ has disappeared from the structure function $\Delta_x(x)$, 
it is easier to analyze the spacetime structure by drawing the function $\Delta_x(x)$ rather than $\Delta_y(y)$. 
For the classification, it is convenient to 
define dimensionless quantities normalized by the acceleration parameter as 
\begin{align}
\label{}
\mathsf Q_0\equiv \sqrt{\frac{A}{|r_0|}}q_0\,, \qquad 
\mathsf G\equiv \frac gA \,,\qquad 
\mathsf R_0 \equiv A r_0 \,.
\end{align}
For $\alpha=1/\sqrt 3$, the last quantity does not appear
in $\Delta_x(x)=1-x^2-\epsilon\mathsf Q_0^2 x^3$, where $\epsilon$ is $+1$ ($-1$) for $r_0>0$ ($r_0<0$). 
Since $\Delta_x(x)$ is a cubic function, it has at least one real root $x_{\ast}$.
It also has two extrema $(x,\,\Delta_x(x))=(0,~1)$ and 
($-2\epsilon/(3\mathsf Q_0^2), 1-4/(27\mathsf Q_0^4)$). 
As the domain with Lorentzian signature, i.e.  $\Delta_x(x)>0$, should be compact, there should be other real roots of $\Delta_x(x)=0$. 
This condition 
 gives rise to the upper bound on the charge as   
 \begin{align}
\label{qrange_m}
\mathsf Q^2_0\leq\frac{2}{3\sqrt 3}\,, ~~~(r_0<0), \\
\mathsf Q^2_0<\frac{2}{3\sqrt 3}\,, ~~~(r_0>0). 
\label{qrange}
\end{align}
It should be noted that the roots become degenerate and the Lorentzian domain disappears when $\mathsf Q^2_0=2/(3\sqrt 3)$ for $r_0>0$.

 The other structure function is now $\Delta_y(x)=\mathsf G^2 -\Delta_x(x)$. 
Thus, the locus of horizons is determined by the intersections of $\mathsf G^2$ and $\Delta_x(x)$.

When $r_0<0$, $\Delta _x(x)$ has two or three real roots $x_-<0<x_+\leq x_{\ast}$ under the condition (\ref{qrange_m}), where  $x_-<x<x_+$ is the Lorentzian domain.
It is easy to find that the intersections take larger values of $x$ than $x_-$ as  $x_-<y_1<y_2<x_+\leq x_{\ast}<y_3$  even if there are three intersections ($\mathsf G^2<1$).
This corresponds to the naked singularity, as we have learned from the 
neutral case in the previous subsection (see figure \ref{fig:PD1}). 
Thus we are compelled to assume $r_0>0$ in the following.

\begin{figure}
\begin{center}
\includegraphics[width=8cm]{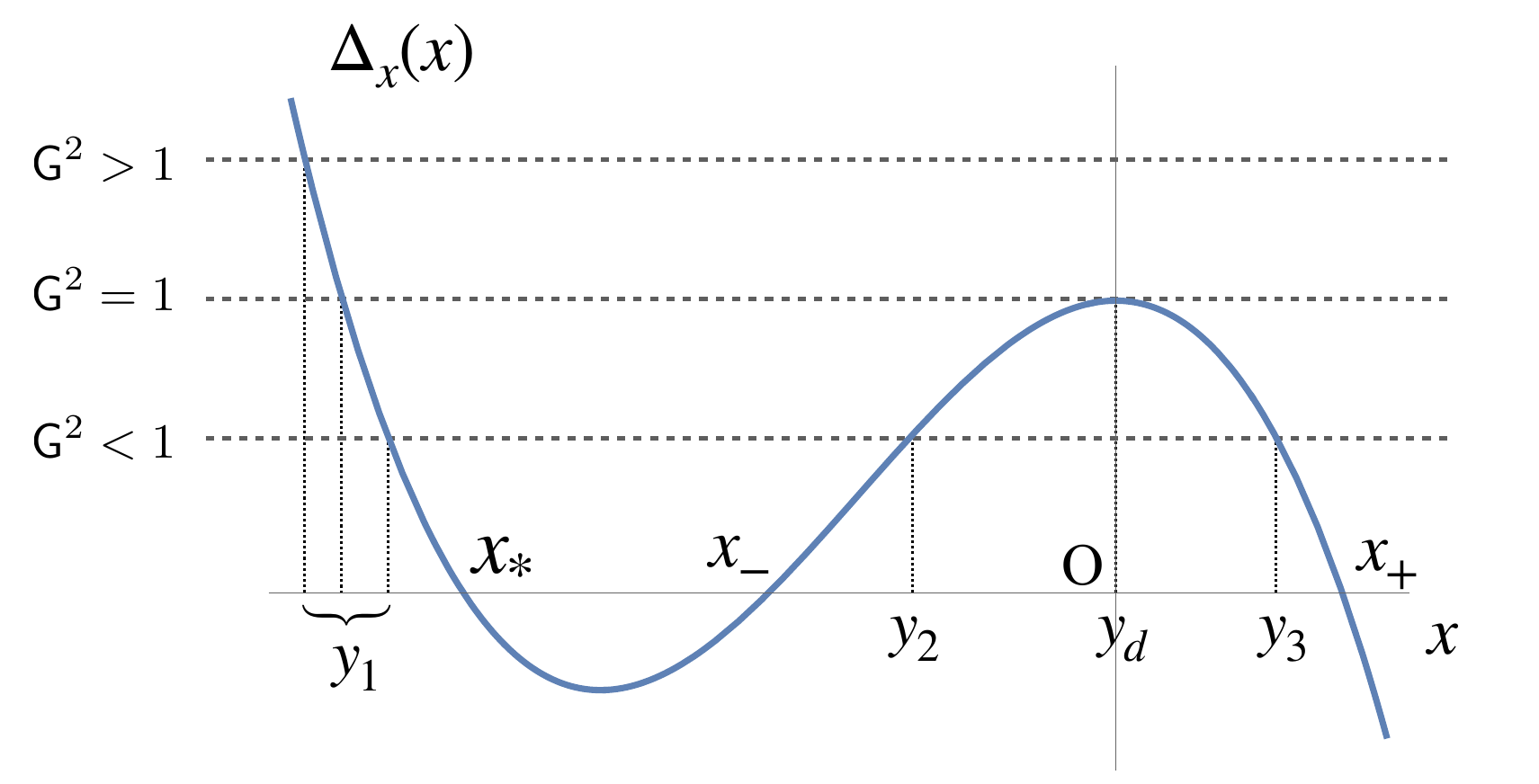}
\caption{Typical behavior of $\Delta_x(x)$ 
in the case where $\Delta_x(x)$ has three distinct real roots $x_*<x_-<x_+$. 
The intersection points of $\Delta_x(x) $ and $\mathsf G^2$ correspond to the Killing horizon. }
\label{fig:CubicDeltax}
\end{center}
\end{figure}

Under the condition  (\ref{qrange}), $\Delta_x(x)=0$ is satisfied at three
distinct real points, which we label $x_{\ast},\, x_-,\, x_+$ in ascending order. 
 The structure function $\Delta_x$ is depicted in figure \ref{fig:CubicDeltax}. 
It is easy to appreciate that these roots obey
\begin{align}
\label{}
-\frac 1{\mathsf Q^2_0}<x_{\ast}<-\frac 2{3\mathsf Q^2_0}<x_-<-1, \qquad 
0<x_+<1 \,. 
\end{align}
Since $\Delta_x(x)$ is a cubic function of $x$, it admits at least one 
 intersection with the line $\mathsf G^2$.
 It gives a real root $y_1$ of $\Delta_y(y)$, which will be identified as an event horizon, and other roots of $\Delta_y(y)$ can be real or complex by the value of  $\mathsf G^2$. 

The configuration of Killing horizons is thus classified into three types 
\begin{itemize}
  \item[(I)] $\mathsf G<1$: three horizons, $y_1<x_-<y_2<y_3 <x_+$. 
  \item[(II)] $\mathsf G=1$: two horizons, $y_1<x_-<y_d <x_+$ where $y_{2}$ and $y_3$ are degenerate $y_2=y_3\equiv y_d \,(=0)$. 
  \item[(III)] $\mathsf G>1$: single horizon, $y_1<x_-<x_+$ where $y_{2}$ and $y_3$ are complex conjugate. 
\end{itemize}
In either case, we have $y_1<x_-$. Hence, the event horizon $y_1$ exists from any angular directions $x\in [x_-, x_+]$.  

\begin{figure}
\begin{center}
\includegraphics[width=15cm]{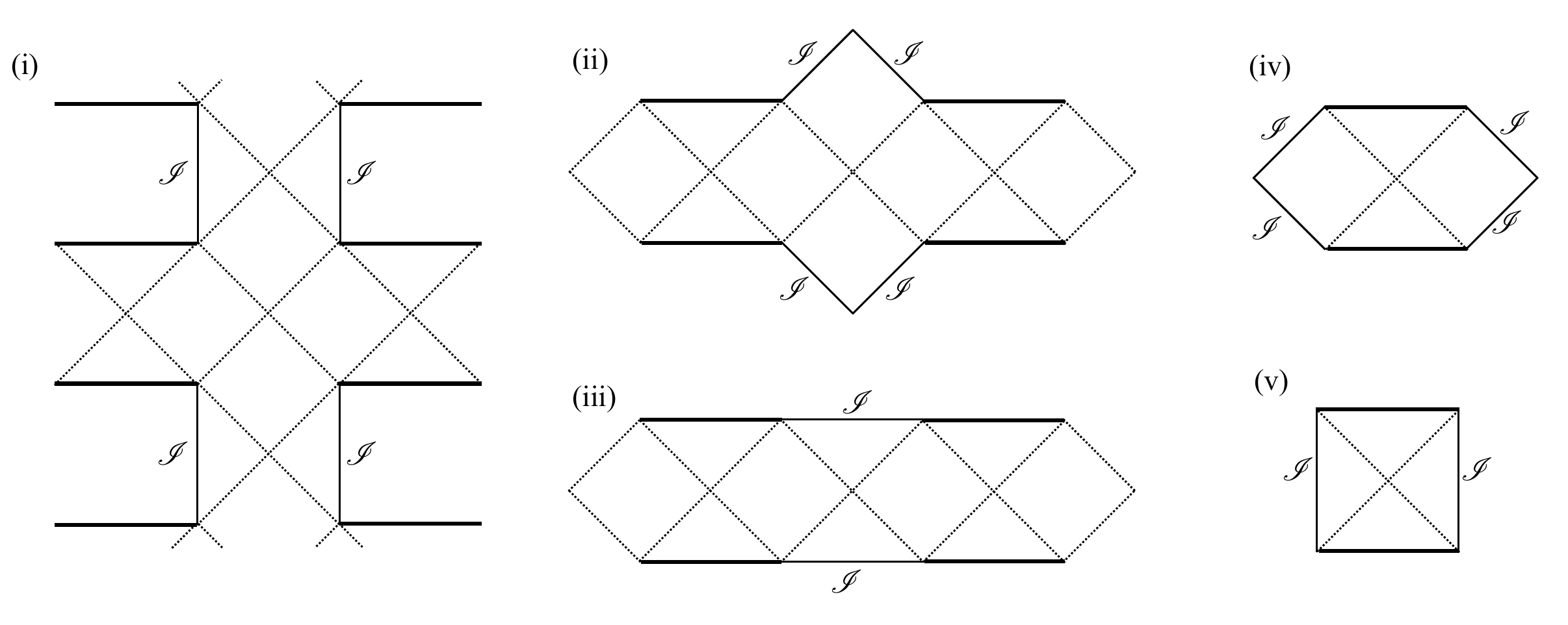}
\caption{Penrose diagrams of the $k=1$ charged C-metric
with $q_1=0$ and $g<A$. Case (i) is for $y_3<x\le x_+$, 
(ii) for $x=y_3$, (iii) for $y_2<x<y_3$, (iv) for $x=y_2$ 
and (v) for $x_-\le x<y_2$. 
The thick lines denote the curvature singularities. }
\label{fig:PDcharged}
\end{center}
\end{figure}

\begin{figure}[t]
\begin{center}
\includegraphics[width=8cm]{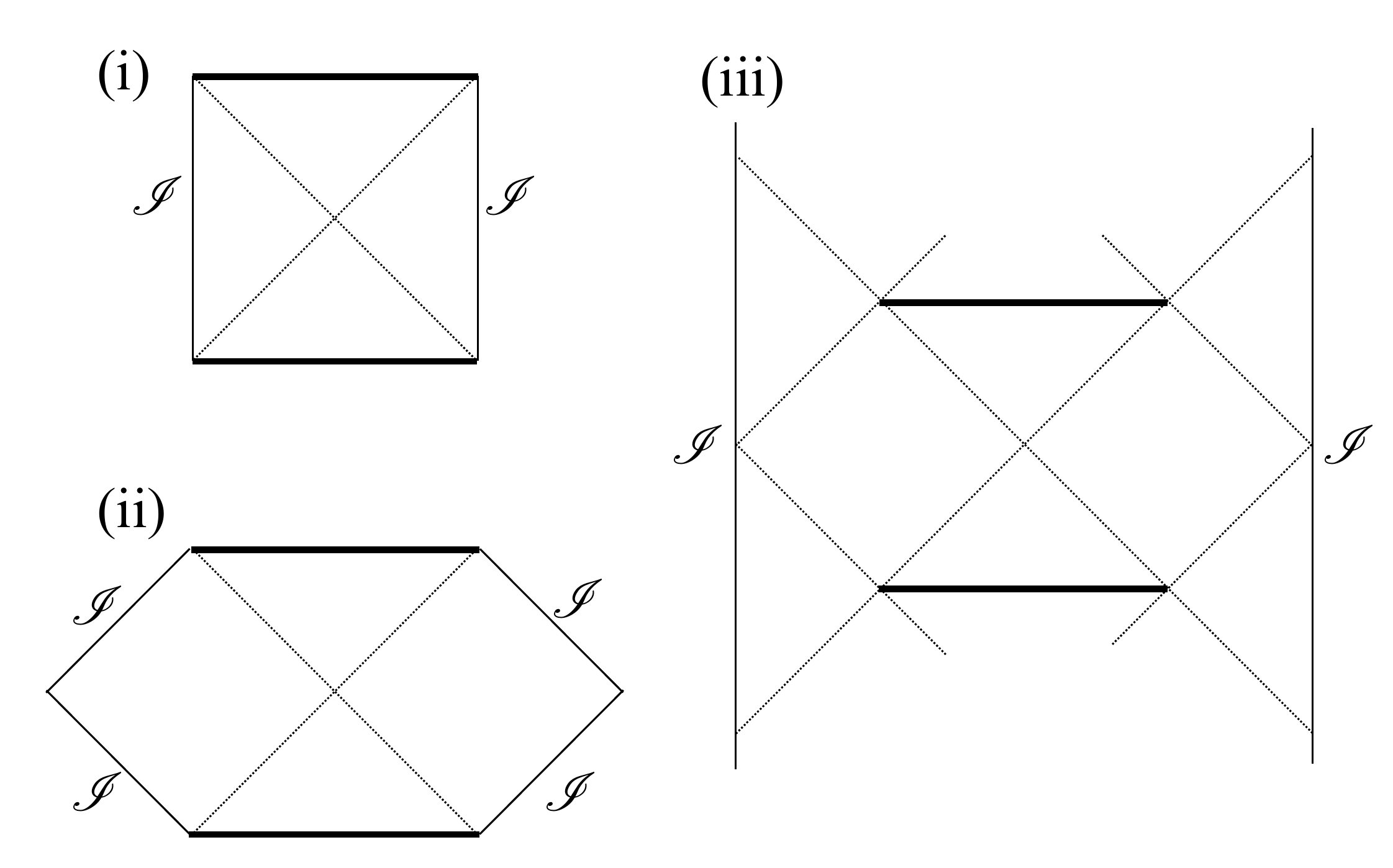}
\caption{Penrose diagrams of the $k=1$ charged C-metric
with $q_1=0$ and $g=A$. Case (i) is for $x_- \le x<y_d=0$, 
(ii) for $x_-=y_d=0$ and (iii) for $0=y_d<x\le x_+$. 
The thick lines denote the curvature singularities. }
\label{fig:PDcharged2}
\end{center}
\end{figure}

Finally we must demand $-1/\mathsf R_0<y_1$ to avoid the naked singularity. 
Since $\Delta_y'(y)=0$ occurs at $y=-2/(3\mathsf Q_0^2)$ and $y=0$,  
this condition amounts to 
$\Delta_y(-1/\mathsf R_0)<0$ and $-1/\mathsf R_0<-2/(3\mathsf Q_0^2)$, 
which gives two kinds of lower bounds on the charge as 
\begin{align}
\label{qrange2}
\frac 23 \mathsf R_0<\mathsf Q_0^2 \,, 
\end{align}
and 
\begin{align}
\label{qrange3}
\mathsf F(\mathsf R_0)<\mathsf Q^2_0\,, ~~~ {\rm where} ~~~
\mathsf F(\mathsf R_0) = \mathsf R_0 \left[1- \mathsf R_0^2 \left(1-\mathsf G^2\right)\right]\,.
\end{align}

For case (I), the condition (\ref{qrange2}) is included in (\ref{qrange3}) under the condition  (\ref{qrange}). 
Hence $\mathsf F(\mathsf R_0)<\mathsf Q^2_0<2/(3\sqrt 3)$ should be satisfied.
The corresponding Penrose diagram is drawn in figure \ref{fig:PDcharged}. 
For case (II), $\mathsf F(\mathsf R_0)=\mathsf R_0>2\mathsf R_0/3$, i.e.,
 $\mathsf R_0<\mathsf Q^2<2/(3\sqrt 3)$ is called for. 
The Penrose diagram is shown in figure \ref{fig:PDcharged2}. 
For case (III), $\mathsf F(\mathsf R_0)$ is monotonic in $\mathsf R_0$ 
with $\mathsf F'(\mathsf R_0)=1>\frac 23$, thereby
we have $\mathsf F(\mathsf R_0)<\mathsf Q^2<2/(3\sqrt 3)$. 
The corresponding Penrose diagram is the same as the 
Schwarzschild-AdS black hole (case (v) of figure \ref{fig:PDcharged}).  

Note that the singularity is always spacelike whenever it is covered by the event horizon, even though the 
solution is charged. 
This is a feature that we have encountered also in a 
 static black hole with a single charge in the asymptotically flat case ($g=0$) \cite{Gibbons:1987ps}.

The cases with the different values $\alpha$ can be analyzed similarly 
despite the augmentation of complexity.  
When either $\Delta_y(y)$ or $h(y)\Delta_y(y)$
is cubic, some numerical tests reveal the need of the relation
\begin{align}
\label{}
-\frac {1}{\mathsf R_0}<y_1<x_-<(y_2\le y_3 <)~x_+ \,. 
\end{align}
We therefore end up with figures \ref{fig:PDcharged}, \ref{fig:PDcharged2} and the Schwarzschild-AdS solution, when 
the curvature singularity is hidden by an event horizon. It would be intriguing to 
extend this analysis for the other values of parameters.

\section{Summary}
\label{summary}

We have constructed a new family of C-metric solutions (\ref{sol1}) in ${\cal N}=2$ gauged supergravity theory with a 
prepotential (\ref{prepotential}). Upon truncation, the theory is nothing but the 
Einstein-Maxwell-dilaton gravity (\ref{Lag}),  in which a real scalar field 
couples to two gauge fields with different coupling constants and 
the scalar potential is expressed in terms of the superpotential. 
Our solution describes a family of C-metrics distinct from the one obtained in the literature \cite{Lu:2014ida,Lu:2014sza}. 
The non-uniqueness of the solutions of this sort has been identified in the vanishing acceleration limit in \cite{Faedo:2015jqa}.
However, the relevance of these solutions remained open to date. We have clarified  in this paper that 
these solutions are converted to each other via the double Wick rotation (\ref{flip}). 
Although these solutions are related in a simple fashion, the physical properties of the solutions 
are considerably different. As we have expounded in section \ref{sec:hairyBH} and \ref{sec:flip},
the most comprehensible  aspect is the presence of the horizons in the uncharged and non-accelerating limit. 
Our solution reduces in the spherical case to the hairy black hole with a regular horizon, whilst the 
the double Wick rotated solution brings about the naked singularity. 

Inspired from the above property, we have investigated the global structure of the neutral solution
in a clear-cut fashion by noting the restriction (\ref{Deltaxy}).
Notably, the neutral solution fails to hide the singularity behind the horizon, as opposed to the non-accelerating case. 
This is our main upshot achieved in this paper. 
A technical crux of this obstruction is that the structure function $\Delta_y(y)$ given by (\ref{Deltays}), which is responsible for the 
horizon structure, is quadratic, while the vacuum structure function $\check \Delta_y(y)$ given by   (\ref{vac:Deltayx}) is cubic. 
The cubic structure can circumvent the thorny constraint (\ref{ypmxpm}). The avoidance of naked singularity
therefore asks for at least one charge. We have verified that this is indeed the case for $\alpha=1/\sqrt 3$. 
Specifically, the charged C-metric is qualified as a pair of accelerated black holes
in AdS, whose causal structure resembles that for the neutral $\Lambda$-vacuum C-metric
\cite{Podolsky:2002nk,Dias:2002mi}.

The present C-metric in supergravity has many potential applications which we set out to delineate in order. 
The $\Lambda$-vacuum C-metric is known to describe an exact black hole in the AdS brane worlds \cite{Emparan:1999wa}. 
Aside from the significance in its own right, these solutions can be applied to explore the 
strongly coupled regime of the boundary conformal field theory \cite{Emparan:2002px} 
and can represent the dual of plasma balls \cite{Emparan:2009dj}. Black funnels and black droplet solutions are also intriguing application \cite{Hubeny:2009kz}. 
To examine the effect of a scalar field would offer a new insight in the holographic context.

The time-dependent generalization of the C-metric is an important testground for the description of 
gravitational radiation. 
This issue has been first discussed in \cite{Gueven:1996zm} within the Robinson-Trautman class \cite{Robinson:1960zzb}, 
which allows a class of non-twisting and shear-free null geodesic congruences. 
 The dynamical generalization of the C-metric considered in \cite{Lu:2014ida} falls into this category. 
The Robinson-Trautman solution outside the Petrov-D class seems an alluring future direction.

A rotating generalization of the electrovacuum C-metric is dubbed as the Pleba\'nski-Demia\'nski solution \cite{Plebanski:1976gy}, which is the 
most general Petrov D solution in Einstein-Maxwell-$\Lambda$ system. 
Unfortunately, rotating solutions within the Einstein-Maxwell-dilaton theory are hard to construct even in the ungauged case, since the target space is not symmetric. A less complicated task is to seek the rotating solution in the original theory (\ref{Lag0}) with a nonvanishing axionic scalar. 
Some solutions for special values of $\alpha$ have been found, but the organizing solution for general $\alpha$ is still missing. 
A promising route is to seek the supersymmetric solutions, for which the possible canonical form of the metric is severely constrained. 
In appendix, we present the conditions under which the C-metric solutions 
(\ref{sol1}) and (\ref{sol2}) preserve supersymmetry. 

The embedding of the Pleba\'nski-Demia\'nski solution into eleven dimensions seems also interesting, since the  uplifted solution can be made regular and is parametrized by quantized conical singularities in four dimensions. 
See e.g. \cite{Ferrero:2020twa,Cassani:2021dwa,Ferrero:2021ovq} for details. 
The obtained $D=11$ solution is holographically interpreted as a membrane wrapped on spindles. For the $\alpha=1, \sqrt 3, 1/\sqrt 3$ cases, 
we expect that the uplifted solution of  our C-metric (\ref{sol1}) can be interpreted in a similar fashion.

Finally, the Euclidean C-metric is a stimulating subject to be explored as well. 
The C-metric instanton solution describes a pair production of black holes by the cosmic string
\cite{Dowker:1993bt,Hawking:1995zn,Eardley:1995au}. Also, the Euclidean dilatonic C-metric 
plays a key role in the construction of vacuum black ring \cite{Emparan:2001wn,Tomizawa:2006vp}. 
Furthermore, the Euclidean Pleba\'nski-Demina\'nski solution enjoys some mathematically 
rich frameworks such as the conformal ambi-K\"ahler structure \cite{Nozawa:2015qea,Nozawa:2017yfl}. 
Pursuing these issues is  left for future investigation.

\acknowledgments
The work of MN is partially supported by MEXT KAKENHI Grant-in-Aid for Transformative Research Areas (A) through the ``Extreme Universe'' collaboration 21H05189 and JSPS Grant-Aid for Scientific Research (17H01091, 20K03929). 
The work of TT is supported by JSPS KAKENHI Grant-Aid for Scientific Research (JP18K03630, JP19H01901) and for Exploratory Research (JP19K21621, JP22K18604).

\appendix

\section{Supersymmetry}

Since the present  Einstein-Maxwell-dilaton theory (\ref{Lag}) originates from the $\mathcal N=2$
supergravity, it is interesting to investigate the supersymmetry preserved by the present solution.
The supersymmetry of the C-metric in the Einstein-Maxwell-$\Lambda$ system has been 
explored in \cite{Klemm:2013eca}. The dilaton coupling constant $\alpha$ induces an interesting effect
on the twist of supersymmetric solutions \cite{Nozawa:2010rf}.

The Killing spinor equations are given by 
\begin{align}
\hat \nabla _\mu \epsilon \equiv \, & \left(\nabla_\mu+\frac{i}{4\sqrt{1+\alpha^2}}\left(e^{-\frac{\alpha\phi}2 }F_{\nu\rho}^{0}
+\alpha e^{\frac{\phi}{2\alpha}} F_{\nu\rho}^{1}\right)\gamma^{\nu\rho}\gamma_\mu +W(\phi)\gamma_\mu -\frac{ig}{\sqrt{1+\alpha^2}} (A^0_\mu+\alpha A^1_\mu) \right)\epsilon=0 \,,\\
 \Pi \epsilon \equiv &\, \left(\gamma^\mu \nabla_\mu \phi -8 \frac{\partial W}{\partial \phi}
+\frac{i}{\sqrt{1+\alpha^2}}\left(-\alpha e^{-\frac{\alpha\phi}2 }F_{\mu\nu}^0+e^{\frac{\phi}{2\alpha}} F_{\nu\rho}^{1}\right)\gamma^{\mu\nu}\right)\epsilon=0  \,,
\end{align}
where $\epsilon $ is a Dirac spinor. 
These equations are derived from the general expressions given in \cite{Cacciatori:2008ek}, 
or fixed by requiring the positive mass theorem \cite{Nozawa:2013maa,Nozawa:2014zia}.

For the existence of nontrivial solutions obeying these first order and algebraic equations, the following integrability
conditions must be fulfilled 
\begin{align}
\label{}
{\rm det}[\hat \nabla _\mu, \hat \nabla _\nu]=0 \,, \qquad 
{\rm det}\,\Pi =0 \,. 
\end{align}
The condition ${\rm det}\,\Pi =0$ for the solution (\ref{sol1}) boils down to 
\begin{align}
\label{BPSsol1}
a_0=-\frac{[(q_0-q_1)^2+a_2 r_0^2]^2}{4A^2q_0^2 r_0^4}\,, \qquad 
a_1=\frac{[(q_0-q_1)(3q_0+q_1)-a_2 r_0^2][(q_0-q_1)^2+a_2 r_0^2]}{8A q_0^2 r_0^3}\,. 
\end{align}
Plugging these results into ${\rm det}[\hat \nabla _\mu, \hat \nabla _\nu]=0$, 
one finds, after lengthy and tedious computations, all the  components of equations are automatically satisfied. 
In this case, the structure function factorizes into 
\begin{align}
\label{}
\Delta_y (y)=\frac{1}{4A^2 q_0^2 r_0^4 h(y)}\left[
-2q_0q_1 +\left\{q_1^2+a_2 r_0^2 +q_0^2(1-2A r_0 y )\right\}h(y)
\right]^2 \ge 0 \,. 
\end{align}
Thus, the Killing horizon for this solution is degenerate, as consistent with supersymmetry.\footnote{
Since the supersymmetry requires the bilinear vector $V^\mu=i\bar \epsilon \gamma^\mu \epsilon$ of the Killing spinor
is a globally defined timelike or null Killing vector, the stationary black-hole horizon, if exists, must be degenerate. If it is nondegenerate, 
the Killing vector $V^\mu$ becomes spacelike inside the black hole.}
In the zero acceleration limit (\ref{para:sol1}), the preservation of supersymmetry requires
\begin{align}
\label{BPSBH1}
q_0 = \pm \left(\frac{k}{2g}+\frac{gr_0^2 \alpha^2(\alpha^2-1)}{(1+\alpha^2)^2}\right)\,, \qquad 
q_1 =  \pm \left(\frac{k}{2g}-\frac{gr_0^2 (\alpha^2-1)}{(1+\alpha^2)^2}\right)\,. 
\end{align}
This occurs for any value of $k$.

For the flipped solution (\ref{sol2}) with (\ref{hatDelta}), the supersymmetric conditions become
\begin{align}
\label{BPSsol2}
a_0=\frac{[(\hat q_0-\hat q_1)^2-a_2 r_0^2]^2}{4A^2\hat q_0^2 r_0^4}\,, \qquad 
a_1=-\frac{[(\hat q_0-\hat q_1)(3\hat q_0+\hat q_1)+a_2 r_0^2][(\hat q_0-\hat q_1)^2-a_2 r_0^2]}{8A \hat q_0^2 r_0^3}\,. 
\end{align}
for which
\begin{align}
\label{}
\hat\Delta_{\hat y} (\hat y)=
\frac{\left[
-2\hat q_0\hat q_1 +\left\{\hat q_1^2-a_2 r_0^2 +\hat q_0^2(1-2A r_0 \hat y )\right\}h(\hat y)
\right]^2}{4A^2 \hat q_0^2 r_0^4 h(\hat y)}
+\frac{g^2}{A^2}h(\hat y)^{\frac{3\alpha^2-1}{1+\alpha^2}} \ge 0\,. 
\end{align}
On account of the fact that $\hat y=-1/(A r_0)$, where $h(\hat y)=0$, is singular, 
it turns out that the flipped solution fails to have a degenerate horizon
in the supersymmetric case. In the zero acceleration limit (\ref{A0metric2}), 
 the supersymmetric condition becomes
\begin{align}
\label{BPSBH2}
(q_0-q_1)^2-k r_0^2=0 \,. 
\end{align}
This equation is satisfied only for $k\ge 0$.

\end{document}